\documentclass[prc,superscriptaddress,showpacs,twocolumn,amssymb,amsmath,amsfonts,aps]{revtex4}
\usepackage[pdftex]{graphicx}
\usepackage{dcolumn}  
\usepackage{longtable}
\usepackage{bm}       
\usepackage{subfigure}
\usepackage{multirow} %
\usepackage{epstopdf}
\newcommand\captionof[1]{\def\@captype{#1}\caption}
\newcommand\myreac{${\gamma p \rightarrow K^{+}\Lambda}$ }

\begin{document}

\newcommand*{\CMU}{Carnegie Mellon University, Pittsburgh, Pennsylvania 15213}
\newcommand*{\CMUindex}{6}
\affiliation{\CMU}
\newcommand*{\WJ}{Washington \& Jefferson College, Washington, PA 15301}
\newcommand*{\WJindex}{1}
\affiliation{\WJ}

\newcommand*{\ANL}{Argonne National Laboratory, Argonne, Illinois 60441}
\newcommand*{\ANLindex}{1}
\affiliation{\ANL}
\newcommand*{\ASU}{Arizona State University, Tempe, Arizona 85287-1504}
\newcommand*{\ASUindex}{2}
\affiliation{\ASU}
\newcommand*{\UCLA}{University of California at Los Angeles, Los Angeles, California  90095-1547}
\newcommand*{\UCLAindex}{3}
\affiliation{\UCLA}
\newcommand*{\CSUDH}{California State University, Dominguez Hills, Carson, CA 90747}
\newcommand*{\CSUDHindex}{4}
\affiliation{\CSUDH}
\newcommand*{\CANISIUS}{Canisius College, Buffalo, NY 14208}
\newcommand*{\CANISIUSindex}{5}
\affiliation{\CANISIUS}
\newcommand*{\CUA}{Catholic University of America, Washington, D.C. 20064}
\newcommand*{\CUAindex}{7}
\affiliation{\CUA}
\newcommand*{\SACLAY}{CEA, Centre de Saclay, Irfu/Service de Physique Nucl\'eaire, 91191 Gif-sur-Yvette, France}
\newcommand*{\SACLAYindex}{8}
\affiliation{\SACLAY}
\newcommand*{\CNU}{Christopher Newport University, Newport News, Virginia 23606}
\newcommand*{\CNUindex}{9}
\affiliation{\CNU}
\newcommand*{\UCONN}{University of Connecticut, Storrs, Connecticut 06269}
\newcommand*{\UCONNindex}{10}
\affiliation{\UCONN}
\newcommand*{\ECOSSEE}{Edinburgh University, Edinburgh EH9 3JZ, United Kingdom}
\newcommand*{\ECOSSEEindex}{11}
\affiliation{\ECOSSEE}
\newcommand*{\FAIRFIELD}{Fairfield University, Fairfield, Connecticut 06824}
\newcommand*{\FAIRFIELDindex}{11}
\affiliation{\FAIRFIELD}
\newcommand*{\FIU}{Florida International University, Miami, Florida 33199}
\newcommand*{\FIUindex}{12}
\affiliation{\FIU}
\newcommand*{\FSU}{Florida State University, Tallahassee, Florida 32306}
\newcommand*{\FSUindex}{13}
\affiliation{\FSU}
\newcommand*{\GWU}{The George Washington University, Washington, DC 20052}
\newcommand*{\GWUindex}{14}
\affiliation{\GWU}
\newcommand*{\ISU}{Idaho State University, Pocatello, Idaho 83209}
\newcommand*{\ISUindex}{15}
\affiliation{\ISU}
\newcommand*{\INFNFR}{INFN, Laboratori Nazionali di Frascati, 00044 Frascati, Italy}
\newcommand*{\INFNFRindex}{16}
\affiliation{\INFNFR}
\newcommand*{\INFNGE}{INFN, Sezione di Genova, 16146 Genova, Italy}
\newcommand*{\INFNGEindex}{17}
\affiliation{\INFNGE}
\newcommand*{\INFNRO}{INFN, Sezione di Roma Tor Vergata, 00133 Rome, Italy}
\newcommand*{\INFNROindex}{18}
\affiliation{\INFNRO}
\newcommand*{\ORSAY}{Institut de Physique Nucl\'eaire ORSAY, Orsay, France}
\newcommand*{\ORSAYindex}{19}
\affiliation{\ORSAY}
\newcommand*{\ITEP}{Institute of Theoretical and Experimental Physics, Moscow, 117259, Russia}
\newcommand*{\ITEPindex}{20}
\affiliation{\ITEP}
\newcommand*{\JMU}{James Madison University, Harrisonburg, Virginia 22807}
\newcommand*{\JMUindex}{21}
\affiliation{\JMU}
\newcommand*{\KNU}{Kyungpook National University, Daegu 702-701, Republic of Korea}
\newcommand*{\KNUindex}{22}
\affiliation{\KNU}
\newcommand*{\LPSC}{LPSC, Universit\'e Joseph Fourier, CNRS/IN2P3, INPG, Grenoble, France
}
\newcommand*{\LPSCindex}{23}
\affiliation{\LPSC}
\newcommand*{\UNH}{University of New Hampshire, Durham, New Hampshire 03824-3568}
\newcommand*{\UNHindex}{24}
\affiliation{\UNH}
\newcommand*{\NSU}{Norfolk State University, Norfolk, Virginia 23504}
\newcommand*{\NSUindex}{25}
\affiliation{\NSU}
\newcommand*{\OHIOU}{Ohio University, Athens, Ohio  45701}
\newcommand*{\OHIOUindex}{26}
\affiliation{\OHIOU}
\newcommand*{\ODU}{Old Dominion University, Norfolk, Virginia 23529}
\newcommand*{\ODUindex}{27}
\affiliation{\ODU}
\newcommand*{\RPI}{Rensselaer Polytechnic Institute, Troy, New York 12180-3590}
\newcommand*{\RPIindex}{28}
\affiliation{\RPI}
\newcommand*{\ROMAII}{Universita' di Roma Tor Vergata, 00133 Rome Italy}
\newcommand*{\ROMAIIindex}{29}
\affiliation{\ROMAII}
\newcommand*{\MSU}{Skobeltsyn Nuclear Physics Institute, Skobeltsyn Nuclear Physics Institute, 119899 Moscow, Russia}
\newcommand*{\MSUindex}{30}
\affiliation{\MSU}
\newcommand*{\SCAROLINA}{University of South Carolina, Columbia, South Carolina 29208}
\newcommand*{\SCAROLINAindex}{31}
\affiliation{\SCAROLINA}
\newcommand*{\JLAB}{Thomas Jefferson National Accelerator Facility, Newport News, Virginia 23606}
\newcommand*{\JLABindex}{32}
\affiliation{\JLAB}
\newcommand*{\UNIONC}{Union College, Schenectady, NY 12308}
\newcommand*{\UNIONCindex}{33}
\affiliation{\UNIONC}
\newcommand*{\UTFSM}{Universidad T\'{e}cnica Federico Santa Mar\'{i}a, Casilla 110-V Valpara\'{i}so, Chile}
\newcommand*{\UTFSMindex}{34}
\affiliation{\UTFSM}
\newcommand*{\ECOSSEG}{University of Glasgow, Glasgow G12 8QQ, United Kingdom}
\newcommand*{\ECOSSEGindex}{35}
\affiliation{\ECOSSEG}
\newcommand*{\VIRGINIA}{University of Virginia, Charlottesville, Virginia 22901}
\newcommand*{\VIRGINIAindex}{36}
\affiliation{\VIRGINIA}
\newcommand*{\WM}{College of William and Mary, Williamsburg, Virginia 23187-8795}
\newcommand*{\WMindex}{37}
\affiliation{\WM}
\newcommand*{\YEREVAN}{Yerevan Physics Institute, 375036 Yerevan, Armenia}
\newcommand*{\YEREVANindex}{38}
\affiliation{\YEREVAN}

\newcommand*{\NOWCUA}{Catholic University of America, Washington, D.C. 20064}
\newcommand*{\NOWJLAB}{Thomas Jefferson National Accelerator Facility, Newport News, Virginia 23606}
\newcommand*{\NOWCNU}{Christopher Newport University, Newport News, Virginia 23606}

\author{M. E. McCracken} 
\affiliation{\CMU}
\affiliation{\WJ}
\author{M. Bellis}
\altaffiliation[Current address:]{Stanford University, Stanford, CA 94305}
\affiliation{\CMU}
\author{C. A. Meyer} 
\affiliation{\CMU}
\author{M. Williams} 
\altaffiliation[Current address:]{Imperial College London, London, SW7 2AZ, UK}
\affiliation{\CMU}

\author {K. P. ~Adhikari} 
\affiliation{\ODU}
\author {M.~Anghinolfi} 
\affiliation{\INFNGE}
\author {J.~Ball} 
\affiliation{\SACLAY}
\author {M.~Battaglieri} 
\affiliation{\INFNGE}
\author {B.L.~Berman} 
\affiliation{\GWU1}
\author{A.S.~Biselli}
\affiliation{\FAIRFIELD}
% ...

\collaboration{The CLAS Collaboration}
\noaffiliation

 \date{\today}

%
 
%
%
%
%
%
%

\date{\today}

\title{Differential cross section 
  and recoil polarization measurements\\
  for the \myreac reaction using CLAS at Jefferson Lab}
%
%
\begin{abstract} 
We present measurements of the differential cross section and $\Lambda$ recoil polarization for the \myreac reaction made using the CLAS detector at Jefferson Lab.
These measurements cover the center-of-mass energy range from 1.62 to 2.84~GeV and a wide range of center-of-mass $K^{+}$ production angles.
Independent analyses were performed using the $K^{+}p\pi^{-}$ and $K^{+}p$ (missing~$\pi^{-}$) final-state topologies; results from these analyses were found to exhibit good agreement.
These differential cross section measurements show excellent agreement with previous CLAS and LEPS results and offer increased precision and a 300~MeV increase in energy coverage.
The recoil polarization data agree well with previous results and offer a large increase in precision and a 500~MeV extension in energy range.
The increased center-of-mass energy range that these data represent will allow for independent study of non-resonant $K^{+}\Lambda$ photoproduction mechanisms at all production angles.
\end{abstract} 
\maketitle

\section{\label{section:intro}Introduction}
The \myreac reaction is a promising channel for the study of excited nucleon resonances.
Because of the pseudoscalar nature of the $K^{+}$ and the self-analyzing decay of the $\Lambda$ baryon, measurement of all polarization observables for this channel is experimentally possible.
Precise measurements of these polarization observables, in addition to the unpolarized differential cross section ($d\sigma/d\cos\theta_{K}^{c.m.}$, where $\theta_{K}^{c.m.}$ is the $K^{+}$ polar angle in the center-of-mass frame), will lead to a full characterization of the channel and an exciting opportunity to assess the contributions of resonant and non-resonant photoproduction mechanisms.
The channel is further simplified by the isospin structure of the final state, which allows coupling only to $I=\frac{1}{2}$ $N^{*}$ intermediate states and not the $I=\frac{3}{2}$ $\Delta^{*}$ states.

Previous large-acceptance measurements of the \myreac differential cross section have been made by the SAPHIR \cite{bockhorst, tran, glander} and CLAS \cite{bradford} collaborations.
The most recent SAPHIR results \cite{glander} are formed from roughly $5.2\times10^{4}$ events and span the center-of-mass energy ($\sqrt{s}$) range from threshold (1.61 GeV) to $\approx2.4$~GeV.
The previous CLAS results draw from approximately $5.6\times10^{5}$ events and represent the $\sqrt{s}$ range from threshold to 2.5~GeV.
Though differing by an order of magnitude in statistics, these results do exhibit troubling discrepancies.
The SAPHIR results are systematically lower ($\approx20\%$) than those of CLAS at forward angles.
While both results show an enhancement in $d\sigma/d\cos\theta_{K}^{c.m.}$ at $\sqrt{s}\approx1.9\textrm{ GeV}$, this enhancement is much more pronounced in the CLAS results (especially for $\cos\theta^{c.m.}_{K}\gtrsim 0$).
Other cross section measurements from LEPS at forward \cite{sumihama} and backward \cite{hicks} angles appear to agree with the CLAS results, but do not overlap with the regions of the CLAS/SAPHIR discrepancy.

These differences have led to difficulties in interpretation of the $N^{*}$ contributions to $K^{+}\Lambda$ production.
Several studies have found evidence for contributions of different resonances dependent upon which results are considered.
The $d\sigma/d\cos\theta_{K}^{c.m.}$ shape discrepancy at $\sqrt{s}\approx1.9\textrm{ GeV}$ is especially problematic, and partial-wave analyses have produced varied explanations for resonant contributions in this region, including $D_{13}$ \cite{mart_bennhold}, $P_{13}$ \cite{shklyar}, $P_{11}$ \cite{sarantsev}, and $S_{11}$~\cite{julia_diaz} states.

In this paper, we present measurements of the ${\gamma p \rightarrow K^{+}\Lambda}$ differential cross section and $\Lambda$ recoil polarization ($P_{\Lambda}$) taken from the CLAS \textit{g11a} dataset.
We have produced separate analyses using the $K^{+}p\pi^{-}$ and $K^{+}p$ (missing $\pi^{-}$) topologies and found these results to be in agreement.
The $K^{+}\Lambda$ measurements presented are the most precise to date, and represent an extension of the observed $\sqrt{s}$ range for $d\sigma /d\cos\theta_{K}^{c.m.}$ and $P_{\Lambda}$ of 300~MeV and 500~MeV, respectively.
These $d\sigma /d\cos\theta_{K}^{c.m.}$ results show agreement with previous CLAS and LEPS results and the $P_{\Lambda}$ results agree well with previous world data.

With several theory groups already pursuing single- and coupled-channel partial-wave analyses including the $\gamma p \rightarrow K^{+}\Lambda$ reaction, the results presented herein will offer new constraints to pre-existing models.
The fine center-of-mass-energy binning of these results, especially the $\Lambda$ recoil polarization, are especially interesting as they show previously unseen structure.
These results also present the first large acceptance measurements of the reaction at center-of-mass energies between 2.53~GeV and 2.84~GeV, an energy regime in which production appears to be dominated by non-resonant processes.
Previous partial-wave analyses have produced non-resonant models by constraining only to forward production angle data or by fitting both resonant and non-resonant components simultaneously.
These $d\sigma/d\cos\theta_{K}^{c.m.}$ and $P_{\Lambda}$ data could allow for independent study of non-resonant production mechanisms at all production angles.

\section{\label{section:exp_setup}Experimental Setup}
Data were collected with the CEBAF Large Acceptance Spectrometer (CLAS) located in Experimental Hall B at the Thomas Jefferson National Accelerator Facility in Newport News, Virginia.
The present results are from the analysis of the CLAS \textit{g11a} dataset, collected during the period of May 17 - July 29, 2004.
Photons were produced via the bremsstrahlung process using a 4.023 GeV electron beam incident on a gold foil.
The Hall B tagger assembly facilitated measurement of the energies of recoil electrons using a dipole magnetic field and scintillator hodoscope; these electron energies were then used to calculate the energy of associated photons \cite{sober}.
After collimation, these photons were incident on the physics target, a cylindrical kapton chamber 40 cm in length and 4 cm in diameter, filled with liquid hydrogen.
Measurements of the target temperature and pressure allowed for calculation of the target density with a relative uncertainty of 0.2\%.

The CLAS detector is composed of tracking and timing detector subsystems arranged with six-fold symmetry about the beamline (\textit{i.e.} in six sectors).
Trajectories of charged particles were deflected by a non-uniform toroidal magnetic field with a maximum magnitude of 1.8 T.
Placement of the physics target allowed for reconstruction of charged tracks leaving the target at polar angles between 8$^{\circ}$ and 140$^{\circ}$.
Charged particle tracking was accomplished with three sets of wire drift chambers per sector.
Event timing information was supplied by the start counter, a thin, segmented scintillation detector placed between the physics target and the innermost tracking components, and the time-of-flight (TOF) wall, a bank of 48 scintillator bars located beyond the outermost tracking component in each sector.
The detector subsystems combined to produce an average relative momentum resolution of approximately 0.5\%.
A more detailed description of the CLAS detector can be found in Ref. \cite{mecking}.

Event triggering required coincident signals from the photon tagger and the CLAS level 1 trigger.
The CLAS level 1 trigger required that two different sectors observe a coincidence between timing signals from the TOF and start counter scintillators.
The signal from the tagger consisted of an OR combination of roughly two-thirds of the tagger's timing scintillators, which corresponded to photons of energy greater than 1.58 GeV.
The timing scintillators corresponding to lower-energy photons were omitted from the trigger in order to reduce the number of recorded events generated by photons below the production threshold for many hadronic final states.
While the number of such events was greatly reduced by this trigger, events generated by photons with energies between $\approx$1.0 GeV and 1.58 GeV could be recorded due to an accidental coincidence with a recoil electron in one of the valid tagger elements.
For the photon spectrum below this energy, a flux renormalization was applied based on the probability of such events.
With this trigger, physics events were recorded and written to disk at a rate of 5 kHz, only a small fraction of which were relevant to this analysis.

\section{\label{section:data}Data and Event Selection}\label{sec:data}
The loose electronics trigger described in the last section allowed for a large number of events to be recorded ($\approx 20\times10^{9}$).
Because only a small fraction of these events were $\gamma p \rightarrow K^{+}\Lambda$ signal, a series of data selection cuts was developed to omit events irrelevant to this analysis (background).

Before physics analysis, the dataset was calibrated.
Timing spectra of the photon tagger, start counter, and TOF subsystems were investigated and corrected.
Drift times from each of the tracking chambers and pulses from TOF scintillators were compared and calibrated.
After these corrections were made, tracks were ``reconstructed'' from raw tracking signals and matched with hits in the start counter and TOF detectors.
Energy and momentum corrections were then applied to individual tracks to account for imperfections in the magnetic field map and detector alignment, and energy losses for particles that traveled through the target, detector material, and air.
Small corrections were also applied to incident photon energies to account for slight deformations in the tagger hodoscope geometry.

As the CLAS detector is optimized for detection of charged particles, only the charged decay mode of the $\Lambda$ ($\Lambda \rightarrow p\pi^{-}$) was considered in this analysis.
Two separate analyses of this reaction were performed: a three-track analysis requiring detection of all three of the final-state particles, and a two-track analysis requiring only the reconstruction of $K^{+}$ and $p$ tracks.
Possible three-track data events were skimmed from the dataset using a 4-constraint kinematic fit to the $\gamma p \rightarrow K^{+}p\pi^{-}$ hypothesis.
Both permutations for the positive track mass hypotheses were tested in all kinematic fits.
This fit imposed energy and momentum conservation by varying the three-momenta of the detected particles within their measurement uncertainties assuming that no undetected particles were involved in the event (missing energy and three-momentum were constrained to zero for a total of four constraints).
A probability that each event came from the desired reaction (confidence level) was then calculated from the variations in momenta and the measurement uncertainties.
For the three-track data, events with confidence levels less than 1\% were removed from the analysis.

Possible two-track data events were selected by performing a 1-constraint kinematic fit to the $\gamma p \rightarrow K^{+}p$ (missing $\pi^{-}$) hypothesis and removing events with confidence levels $<5\%$.
Because the $\pi^{-}$ was not reconstructed, this fit imposed only a single constraint that the missing mass be that of a $\pi^{-}$.
In order to produce results for which uncertainties are dominated by systematic rather than statistical uncertainties, it was sufficient to analyze only 28\% of the full dataset in producing the two-track sample. 
Both data samples were then separated into 10-MeV-wide $\sqrt{s}$ bins.
The uncertainty in the resulting differential cross section measurements due to differences in signal lost to the confidence level cuts in data and Monte Carlo was estimated to be 3\% \cite{omega_paper}.

\section{\label{section:intro}Background Reduction}
Different methods for background subtraction were developed for the two- and three-track analyses.
The skim described in the previous section used a kinematic fit, considering tracks' four-momenta and detector resolution to select event candidates for the $\gamma p \rightarrow pK^{+}\pi^{-}$ reaction.
Particle identification was then refined by considering timing information for each of the positively charged final-state tracks.

\begin{figure*}[]
  \centering
  \rotatebox{90}{\includegraphics[width=0.5\textwidth]{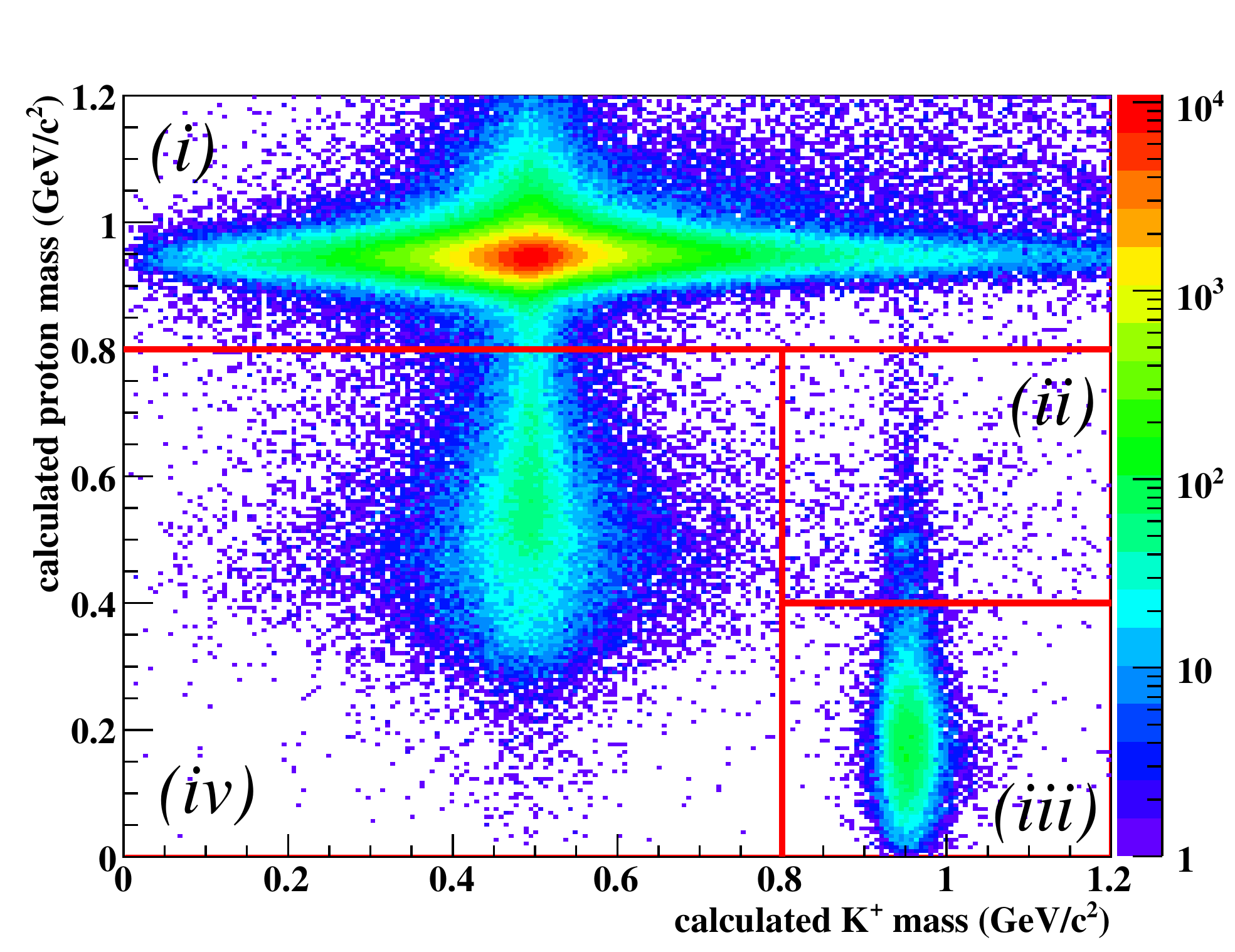}}
  \caption[]{\label{fig:3t_calcmass}
  (Color On-line)  
calculated mass of $p$ tracks \textit{vs.} calculated mass of $K^{+}$ tracks for all events in the three-track data sample: Regions $(i)$ and $(iv)$ correspond to $\gamma p \rightarrow K^{+}\Lambda$ signal events, while regions $(ii)$ and $(iii)$ represent background and are cut from the analysis.
  See text for details.
}
\end{figure*}

For the three-track analysis, the post-kinematic-fit data sample was relatively free of background; background events comprised less than $2.5\%$ of the sample for all values of center-of-mass energy, $\sqrt{s}$.
To further investigate the nature of this sample, the calculated mass, $m_{c}$, was constructed for each track according to
\begin{equation}
m_{c} = \sqrt{\vec{p}^{2}(1-\beta^{2})/\beta^{2}c^{2}},
\end{equation}
where $\vec{p}$ and $\beta$ are the momentum and velocity for the particle as calculated from tracking and timing information.
By considering two-dimensional histograms of the calculated masses of the hypothesized proton and $K^{+}$ tracks, the nature of the remaining background is discernible (see Fig. \ref{fig:3t_calcmass}).
Region $(i)$ in Fig. \ref{fig:3t_calcmass} contains events for which the tracks have calculated masses appropriate of the $p$ and $K^{+}$, thus identifying it as a signal-rich region.
Region $(iii)$ contains events for which the $K^{+}$ track is actually a misidentified proton and the proton track is a misidentified $\pi^{+}$ (\textit{i.e.}, the event is a misidentified $p\pi^{+}\pi^{-}$ final state).
The majority of the background events lie in this region.
Region $(ii)$ represents events that passed the kinematic fit with the proton and $K^{+}$ tracks reversed (\textit{i.e.}, $K^{+}$ misidentified as a proton, proton misidentified as a $K^{+}$).
Events that populate this region are also present in region $(i)$ with the correct identification of $p$ and $K^{+}$.
Region $(iv)$ contains events for which the proton track does not appear to have an appropriate calculated mass; however, further investigation of these events reveals that they are $\gamma p \rightarrow K^{+}p\pi^{-}$ events for which the proton timing information was distorted by the detector.
(This effect is also present in the Monte Carlo, so this small fraction of signal events was not removed from the analysis.)
To remove events from regions $(ii)$ and $(iii)$ from the analysis, a loose two-dimensional cut on the calculated masses for the $K^{+}$ and proton tracks was used, requiring $m_{c}(p)>0.800$~GeV OR $m_{c}(K^{+})<0.800$~GeV.

In each $\sqrt{s}$ bin, we fit a Gaussian function to the missing mass off of the $K^{+}$ distribution and removed any events for which this quantity was greater than 2.5 standard deviations from the mean.
Finally, fiducial cuts were applied to remove events from kinematic or detector regions that could not be reliably modeled.
The three-track data sample included $\approx1.5\times10^{6}$ signal events occupying the $\sqrt{s}$ range from 1.63 to 2.84~GeV, which analysis and fiducial cuts reduced to $\approx 6.5\times10^{5}$ events with less than 1\% background content at all $\sqrt{s}$.
Signal loss to particle identification cuts for this topology was found to be less than $0.11\%$.

Because of the less-restrictive kinematic fit for the two-track analysis, this data sample had a larger percentage of background events.
To mitigate this, we first applied the same cut on the calculated proton and $K^{+}$ masses described above.
For $\sqrt{s}\leq1.660$~GeV, an additional two-dimensional calculated mass cut was used to remove $p\pi^{+}$ background.
This cut kept events for which
\begin{equation}
  m_{c}(p) < m_{c}(K^{+}) + 0.75\textrm{ GeV}/c^{2}.
\end{equation}
These cuts remove roughly half of the background events, and the Feldman-Cousins method \cite{feldman-cousins} was used to estimate signal loss to be less than $0.45\%$ for $\sqrt{s}\leq 1.66$ GeV and less than $3.4\%$ for $\sqrt{s}>1.66$ GeV.
Fiducial cuts were applied as in the three-track analysis.

\begin{figure*}[]
  \centering
  \subfigure[]{
    \rotatebox{90}{\includegraphics[width=0.35\textwidth]{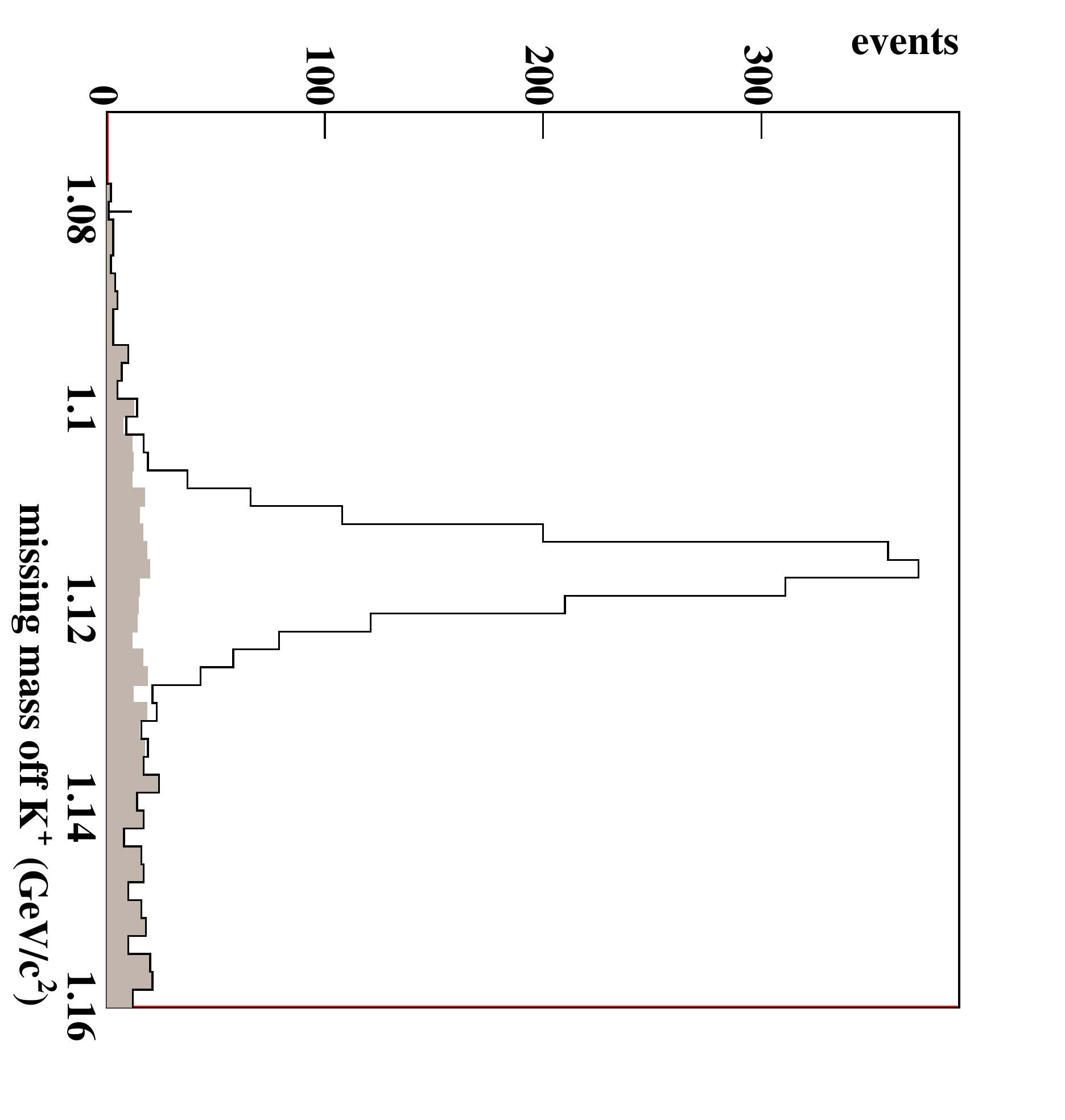}}
  }
  \subfigure[]{
    \rotatebox{90}{\includegraphics[width=0.35\textwidth]{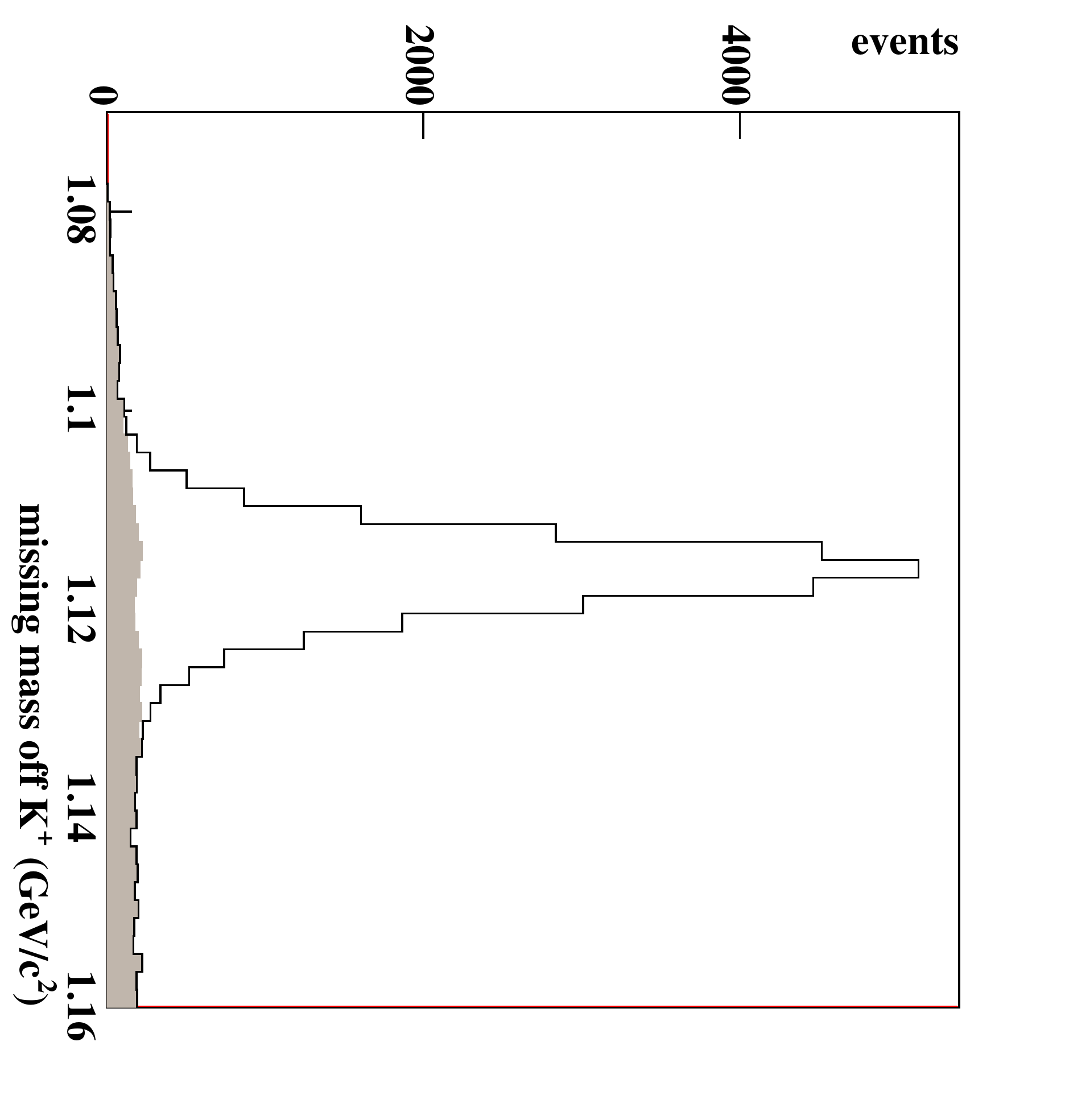}}
  }
\caption[]{\label{fig:sig_bkgd}
  The histograms above show the missing mass off $K^{+}$ distributions for events in the $\sqrt{s}=2.105$~GeV bin of the two-track data sample.
  Figure (a) shows the distribution for events with $-0.05\leq\cos\theta_{K}^{c.m.}<0.05$.
  The unshaded histogram shows all events, whereas the shaded histogram shows the same events weighted by $(1-Q_{i})$.
  Figure (b) shows all events in the $\sqrt{s}=2.105$~GeV bin (no restriction on $\cos\theta_{K}^{c.m.}$) with the same shading scheme.
  See text for details.
}
\end{figure*}

We then applied to the two-track sample an event-based background subtraction technique described in Ref. \cite{bkgd_sub}.
This procedure assigns to each event a quality factor ($Q$-factor) that was used to weight the event's contribution to the fit and the differential cross section calculation.
We defined a metric based upon the cosine of the $K^{+}$ production angle in the c.m. frame ($\cos\theta_{K}^{c.m.}$) and cosine of the proton momentum polar angle ($\cos\theta^{\Lambda HF}_{p}$) and proton azimuthal angle ($\phi^{\Lambda HF}_{p}$) in the $\Lambda$ helicity frame.
For a given event $i$, this metric was then used to identify event $i$'s 100 ``nearest neighbors.''
An unbinned maximum-likelihood fit of a Gaussian signal ($s(m)$) and linear background ($b(m)$) functions was then performed to the missing mass off $K^{+}$ values for these 101 events.
For event $i$, the $Q$-factor was then calculated from the signal and background functions:
\begin{equation}
  Q_{i} = s_{i}/(s_{i}+b_{i}).
\end{equation}
An example of the signal and background separation in a single $\sqrt{s}$ bin is shown in Fig.~\ref{fig:sig_bkgd}.
We then summed the $Q$-factors for all events to estimate the number of signal events present in the two-track data sample after all cuts to be $\approx 1.66\times10^{6}$. 

\section{\label{section:acc}Detector Acceptance}
Monte Carlo modeling of the detector acceptance was done using GSIM, a GEANT-based  simulation of the CLAS detector.
$3\times10^{8}$ $\gamma p \rightarrow K^{+}\Lambda$ events were pseudo-randomly generated according to a phase-space distribution.
GSIM was used to simulate the detector's effects on these ``raw'' events, and a set of ``accepted'' Monte Carlo events was obtained after processing.
(GEANT was also used to simulate the $\Lambda \rightarrow p \pi^{-}$ decay, assuming no net polarization for the hyperons.)
Corrections accounting for the efficiency of the event trigger were applied based on efficiencies of individual timing components (TOF and start counter).
Accepted Monte Carlo events were processed with the same series of analysis cuts as the data events.
An additional momentum smearing algorithm was used to match the momentum resolution of the accepted Monte Carlo events with that of the data. 
More detailed descriptions of the full detector simulation can be found in Ref. \cite{thesis}.

In order to form an accurate characterization of CLAS's acceptance for the $\gamma p \rightarrow K^{+}\Lambda$ reaction, the accepted Monte Carlo events were weighted to resemble the data following the work in Ref. \cite{omega_paper}.
To do this, we expanded the scattering amplitude, $\mathcal{M}$, for the reaction in a large set of basis states:
\begin{equation}\label{eq:scat_exp}
\mathcal{M}_{m_{\gamma},m_{i},m_{\Lambda}}(\vec{X},\vec{\alpha}) \approx \displaystyle \sum_{J=\frac{1}{2}}^{\frac{11}{2}} \sum_{P=\pm} \mathcal{A}^{J^{P}}_{m_{\gamma},m_{i},m_{\Lambda}}(\vec{X},\vec{\alpha}),
\end{equation}
where $m_{\gamma}$, $m_{i}$, and $m_{\Lambda}$ are the spin projections along the beam direction of the incident photon, target proton, and $\Lambda$, respectively; $\vec{X}$ represents the physically significant kinematic quantities ($\cos\theta_{K}^{c.m.}$, $\cos\theta^{\Lambda HF}_{p}$ and $\phi^{\Lambda HF}_{p}$); $\mathcal{A}$ are the $s$-channel partial-wave amplitudes for an intermediate spin-parity $J^{P}$ state (using $\frac{1}{2} \leq J\leq \frac{11}{2}$ and $P=\pm$); and $\vec{\alpha}$ denotes a set of 34 fit parameters.
For this expansion, the $s$-channel partial-wave amplitudes for the $\gamma p \rightarrow K^{+}\Lambda$ reaction serve as basis states and were calculated for each data and Monte Carlo event using the {\tt qft++} package \cite{qftpp}.
Estimators, $\hat{\alpha}$, for the fit parameters were then obtained via unbinned maximum-likelihood fits to the data in each $\sqrt{s}$ bin.
We stress that the results of this fit are not interpreted as physically meaningful (\textit{i.e.}, they do not describe resonant contributions to the reaction); the fit results merely express the expansion scattering amplitude prescribed by the data.

\begin{figure}[]
  \centering
  \rotatebox{90}{\includegraphics[width=0.45\textwidth]{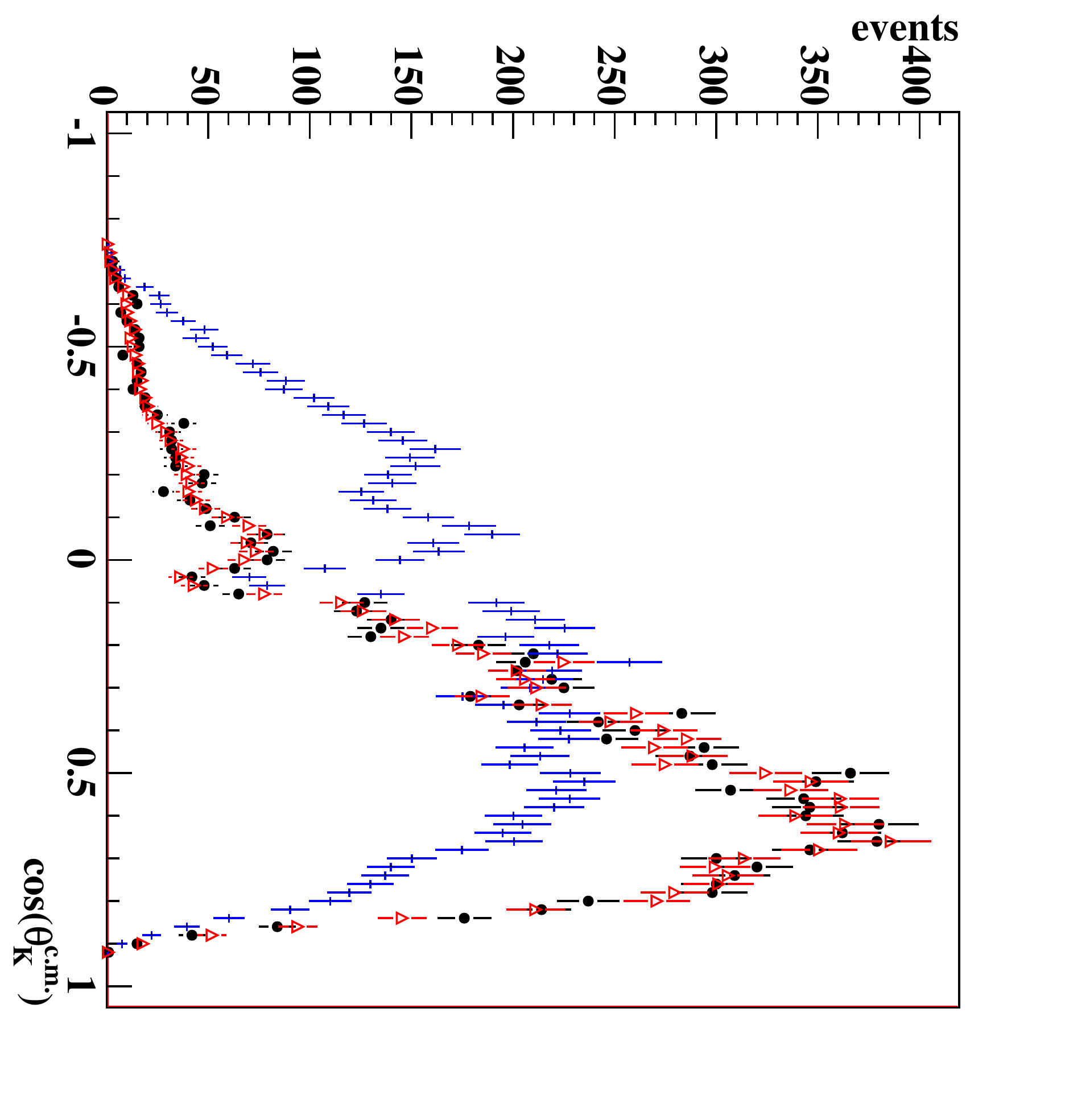}}
\caption[]{\label{fig:weighted_acc}
  (Color On-line)
  Shown above are $\cos\theta_{K}^{c.m.}$ distributions for events in the $\sqrt{s}=2.105$~GeV bin.
  The data and unweighted accepted Monte Carlo are shown by black circles and blue crosses, respectively.
  The Monte Carlo distribution weighted according to the scattering amplitude expansion is shown by red triangles; this weighted distribution matches that of the data events.
  See text for details.
}
\end{figure}

Based on this expansion of the data, we then assigned to each Monte Carlo event $i$ a weight, $I_{i}$, given by
\begin{equation}
  I_{i} = \displaystyle \sum_{m_{\gamma},m_{i},m_{\Lambda}}|\mathcal{M}_{m_{\gamma},m_{i},m_{\Lambda}}(\vec{X}_{i},\hat{\alpha})|^{2}.
\end{equation}
The weighted accepted Monte Carlo matches the data in distributions of all physically significant observables and their correlations (See Fig.~\ref{fig:weighted_acc}), indicating that our set of basis states is large enough to insure a good fit.
We then calculate the detector acceptance, $\eta$, for a region described by kinematic variables $\vec{X}$ as
\begin{equation}
  \eta(\vec{X}) = \left( \displaystyle \sum_{i}^{N_{acc}} I_{i} \right) / \left( \displaystyle \sum_{j}^{N_{raw}}I_{j}\right),
\end{equation}
where the numerator and denominator sums are over the accepted and raw Monte Carlo events, respectively.

It should be noted that this expansion method of acceptance calculation replaces the method of attempting to generate a physics model for Monte Carlo generation via iteration, a method that can be complicated if background is present in the data.
This expansion method also allows for separate calculation of the acceptance for each of the six sectors of CLAS.  
To estimate systematic uncertainties in this acceptance calculation, acceptance-corrected data yields were calculated independently for each sector of CLAS in twelve $\sqrt{s}$ bins.
By considering the variation in these acceptance-corrected yields in each bin, a $\sqrt{s}$-dependent uncertainty ($\sigma_{\eta}(\sqrt{s})$) in the acceptance calculation was determined to be
\begin{equation}
\sigma_{\eta}(\sqrt{s}) = 0.0243\sqrt{s}/\textrm{GeV} - 0.00890,
\end{equation}
which ranges from 3.0\% at threshhold to 6.0\% at ${\sqrt{s} = 2.835}$~GeV.

\section{\label{section:rec_pol}Extracting $P_{\Lambda}$}
The expansion of the data described in Eq. (\ref{eq:scat_exp}) allows for an elegant and efficient extraction of the $\Lambda$ recoil polarization (similar to that described in Ref. \cite{omega_paper}).
In this expansion, we have chosen to represent the photon momentum as the $\hat{z}$ direction, choosing the remaining axes such that the transverse $K^{+}$ momentum is parallel to $\hat{x}$ and $\hat{y}$ extends perpendicular to the reaction plane. 
The $\Lambda$ recoil polarization, $P_{\Lambda}$, is a measure of the $\Lambda$ baryon's polarization out of the reaction plane, \textit{i.e.} along the $\hat{y}$-axis.
As we have written our amplitudes in terms of the $\hat{z}$-projections of the photon, target proton, and $\Lambda$ spins, the recoil polarization at a given value of the kinematic variables, $\vec{X}$, can be easily projected from the scattering amplitude.

To do so, we first construct a two-component wave function, $\psi$, given by
\begin{equation}
\psi(\vec{X}) = \left(\begin{array}{c}
    A^{m_{\gamma},m_{i},M=+}(\vec{X})\\
    A^{m_{\gamma},m_{i},M=-}(\vec{X}) \end{array}\right),
\end{equation}
where $M=\pm$ indicates the spin projection of the $\Lambda$ along the $z$-axis and $A$ is the scattering amplitude evaluated for the appropriate spin projections and kinematics.
$P_{\Lambda}$ is then projected with a simple application of $\sigma_{y}$, the Pauli spin matrix in the $S_{z}$-basis:
\begin{eqnarray}
  P_{\Lambda} &=& \frac{1}{N}\displaystyle \sum_{m_{\gamma},m_{i}}\psi^{\dagger}\sigma_{y}\psi\\
  &=& \frac{i}{N}\displaystyle \sum_{m_{\gamma},m_{i}}(A^{m_{\gamma},m_{i},+}A^{*m_{\gamma},m_{i},-} \nonumber\\
  & & \;\;\;\;\;\;\;\;\;\;\;\;\;\;\;\;- A^{*m_{\gamma},m_{i},+}A^{m_{\gamma},m_{i},-}),
\end{eqnarray}
where
\begin{equation}
N = \displaystyle\sum_{m_{\gamma},m_{i}}\displaystyle\sum_{M}|A^{m_{\gamma},m_{i},M}(\vec{X})|^{2}
\end{equation}
is a normalization factor.

This projection method presents several benefits over traditional methods of fitting proton asymmetry distributions.
Because expansion of the scattering amplitude is used for the acceptance calculation method, extraction of $P_{\Lambda}$ from amplitudes requires little further analysis.
The traditional method requires independent fits to proton momentum asymmetry distributions in a large number of $(\sqrt{s},\cos\theta_{K}^{c.m.})$ bins.
This compound binning of the data can lead to low statistics in some kinematic bins, making binned $\chi^{2}$ fits in these bins difficult to interpret due to large parameter uncertainties.
The method presented here is both more efficient, requiring only a single global fit in each $\sqrt{s}$ bin (see Sec. \ref{section:acc}), and more stable (with respect to iterations and initial parameter values) due to the use of unbinned maximum-likelihood fitting.
Lastly, any $P_{\Lambda}$ measurements given by this method are constrained to lie within the physical range, \textit{i.e.} $P_{\Lambda}\in [1,-1]$, a feature that is not guaranteed by the traditional extraction method.

\section{\label{section:norm}Normalization}
The photon flux during this experiment's run period was determined by measuring the rate for electrons incident on the photon tagger not corresponding to a triggered physics event in CLAS.
Corrections were made to account for the live time of the data acquisition system.
Photon attenuation between tagger and physics target was studied using a total absorption counter downstream of CLAS.
Taking these effects into consideration, an energy-dependent total photon flux was calculated according to the energy segmentation of the tagger hodoscope.
More information on the flux normalization calculation can be found in Ref.~\cite{eta_paper}.

Faulty tagger electronics prevented accurate electron rate measurement for photons in the energy range $2.730\textrm{ GeV}\leq\sqrt{s}<2.750\textrm{ GeV}$.
Intricacies of the event trigger also prevented an accurate flux calculation for the $\sqrt{s}$ bin at 1.955~GeV.
Events in this $\sqrt{s}$ bin could be catalyzed by photons corresponding to both the triggered and un-triggered regions of the tagger.
Thus, we present no differential cross section results for the $\sqrt{s} =$1.955, 2.735, and 2.745~GeV bins.
However, as recoil polarization measurements do not depend on the photon flux, we do present $P_{\Lambda}$ measurements at these energies.

\section{\label{section:syst}Systematic Uncertainties}
By considering acceptance-corrected yields from individual sectors of CLAS, we have estimated a $\sqrt{s}$-dependent acceptance uncertainty between 3\% and 6\% (see Sec. \ref{section:acc}).
Uncertainties due to signal loss to particle identification cuts have been estimated to be 0.1\% for the three-track topology and 0.5\% and 3\% for the two-track topology for bins with $\sqrt{s}>1.660$~GeV and $\sqrt{s}\leq1.660$~GeV, respectively.
Uncertainty due to kinematic fit confidence level cuts has been estimated to be 3\% using a study of confidence level distributions for the $\gamma p \rightarrow p \pi^{+}\pi^{-}$ reaction in this same dataset \cite{omega_paper}.
A 0.5\% uncertaintly for the $\Lambda\rightarrow p \pi^{-}$ branching fraction has been included.
Uncertainty in the target length and fluctuations in its density contribute 0.2\%.
Uncertainty in the flux calculation for this dataset, including effects of photon transmission efficiency and live-time calculations, has been estimated to be 8\% \cite{omega_paper}.
Systematic uncertainties as they contribute to the two- and three-track $d\sigma/d\cos\theta_{K}^{c.m.}$ measurements are outlined in Table \ref{tab:systs}.
These individual uncertainties are combined in quadrature to yield an overall systematic uncertainty for $d\sigma/d\cos\theta_{K}^{c.m.}$ measurements of 9\%-11\%, dependent on topology and center-of-mass energy.

\begin{table*}[]
  \centering
  \begin{tabular}{|l|r|r|r|}
    \hline
    \multirow{2}{*}{Error} & \multirow{2}{*}{$pK^{+}\pi^{-}$} & \multicolumn{2}{|c|}{$\;\;\;\;pK^{+}(\pi^{-})$} \\ 
    \cline{3-4}
    & & $\sqrt{s}<1.66$ GeV & $\sqrt{s}\geq1.66$ GeV \\ \hline
    Particle ID & 0.1\% & 3\% & 0.5\% \\ 
    Confidence Level Cuts & 3\% & 3\% & 3\% \\ 
    Acceptance & 3\%-6\% & 3\%-6\% & 3\%-6\% \\ 
    Normalization & 8\% & 8\% & 8\% \\
    Target Characteristics & 0.2\% & 0.2\% & 0.2\% \\
    $\Lambda \rightarrow p \pi^{-}$ Branching Fraction & 0.5\%  & 0.5\%  & 0.5\% \\ \hline
    Total & 9\%-10.4\% & 10\%-11\% & 9\%-10.4\% \\
    \hline
    \end{tabular}
  \caption[]{\label{tab:systs} Systematic uncertainties in this analysis.  The added particle identification cut applied to the two-track analysis at low $\sqrt{s}$ leads to a larger uncertainty.  These bins are treated separately.
}
\end{table*}

Because measurement of the $\Lambda$ recoil polarization does not depend on target characteristics or flux normalization, uncertainties associated with these factors do not contribute to the uncertainty in $P_{\Lambda}$.
Our $P_{\Lambda}$ extraction method provides no \textit{a priori} method for calculating the associated systematic uncertainty.
The effect of acceptance uncertainty on $P_{\Lambda}$ has been studied by considering results given by alternate acceptance scenarios \cite{thesis} and has been estimated to be 0.05 for both two- and three-track topologies.

\section{\label{section:results}Results}
\subsection{Differential Cross Section}

\begin{figure*}[p]
  \clearpage
  \centering
\hspace{-1.15cm}\rotatebox{90}{\hspace{8.5cm}\Large{$\frac{d\sigma}{d\cos\theta_{K}^{c.m.}}$ ($\mu$b)}}\includegraphics[width=6.5in]{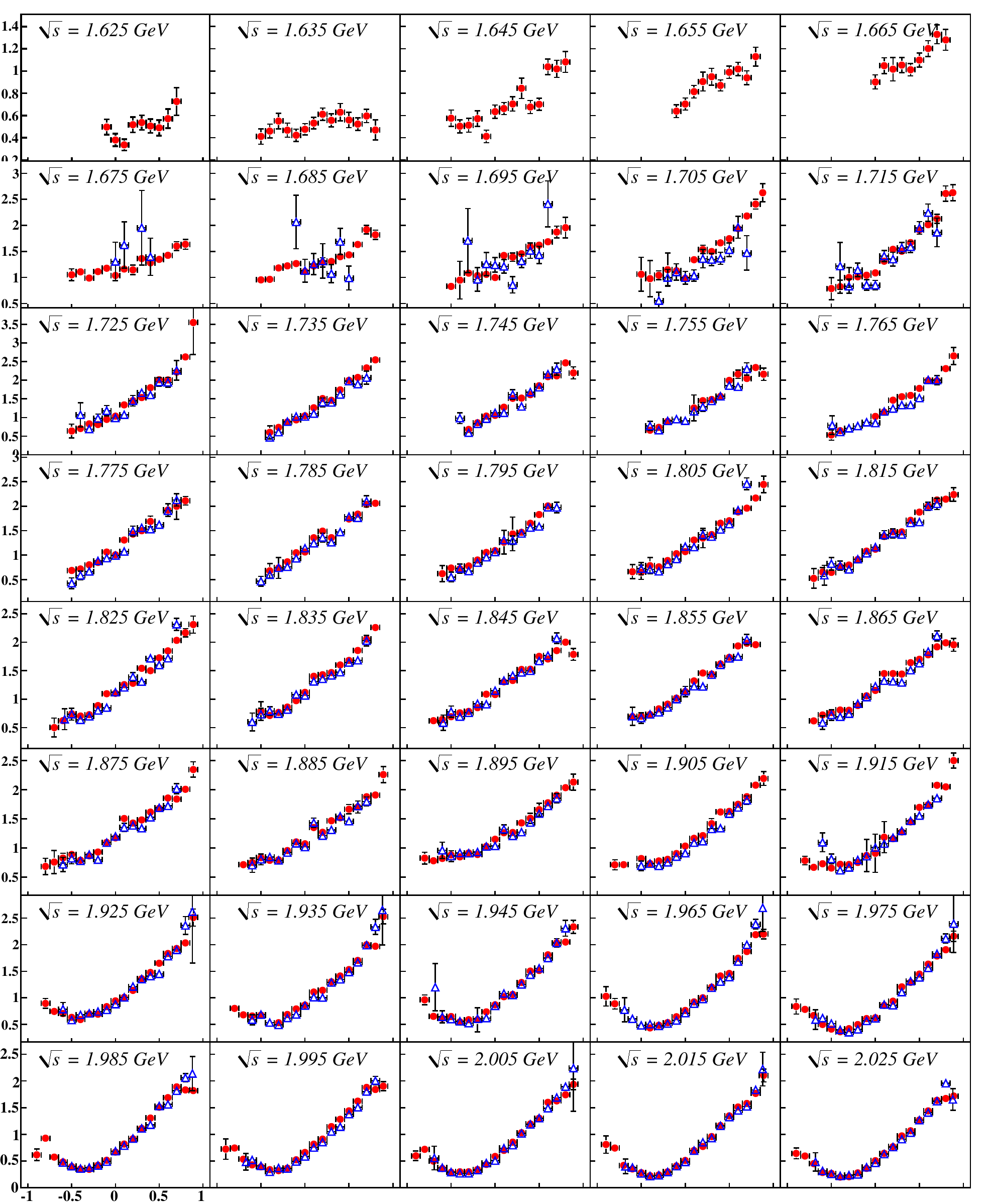}\\
\Large{$\cos\theta_{K}^{c.m.}$}
\caption[]{\label{fig:dsig0}
  (Color On-line)
  $\frac{d\sigma}{d\cos\theta_{K}^{c.m.}}$ ($\mu$b) {\em vs.} $\cos\theta_{K}^{c.m.}$  in bins of $\sqrt{s}$.
Results from the two-track analysis are represented by closed red circles, those of the three-track analysis by open blue triangles.  All error bars represent statistical uncertainties only.  Vertical axes have the same scale in each row, and horizontal axes all have the same scale.
}
\end{figure*}

\begin{figure*}[p]
  \clearpage
  \centering
\hspace{-1.15cm}\rotatebox{90}{\hspace{8.5cm}\Large{$\frac{d\sigma}{d\cos\theta_{K}^{c.m.}}$ ($\mu$b)}}\includegraphics[width=6.5in]{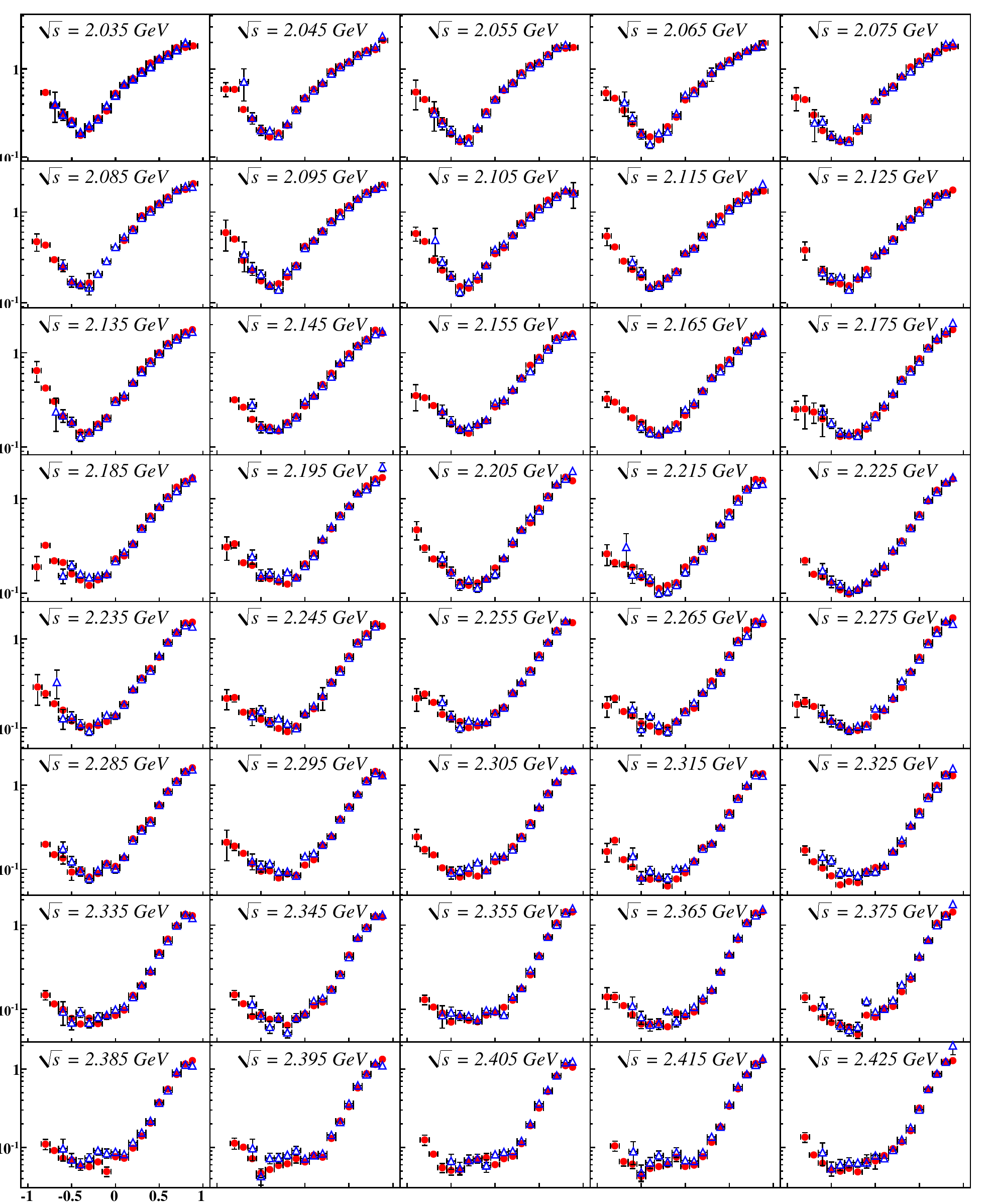}\\
\Large{$\cos\theta_{K}^{c.m.}$}
\caption[]{\label{fig:dsig1}
  (Color On-line)
  $\frac{d\sigma}{d\cos\theta_{K}^{c.m.}}$ ($\mu$b) {\em vs.} $\cos\theta_{K}^{c.m.}$  in bins of $\sqrt{s}$.
Results from the two-track analysis are represented by closed red circles, those of the three-track analysis by open blue triangles.  All error bars represent statistical uncertainties only.  Vertical axes have the same scale in each row, and horizontal axes all have the same scale.
}
\end{figure*}

\begin{figure*}[p]
  \clearpage
  \centering
\hspace{-1.15cm}\rotatebox{90}{\hspace{8.5cm}\Large{$\frac{d\sigma}{d\cos\theta_{K}^{c.m.}}$ ($\mu$b)}}\includegraphics[width=6.5in]{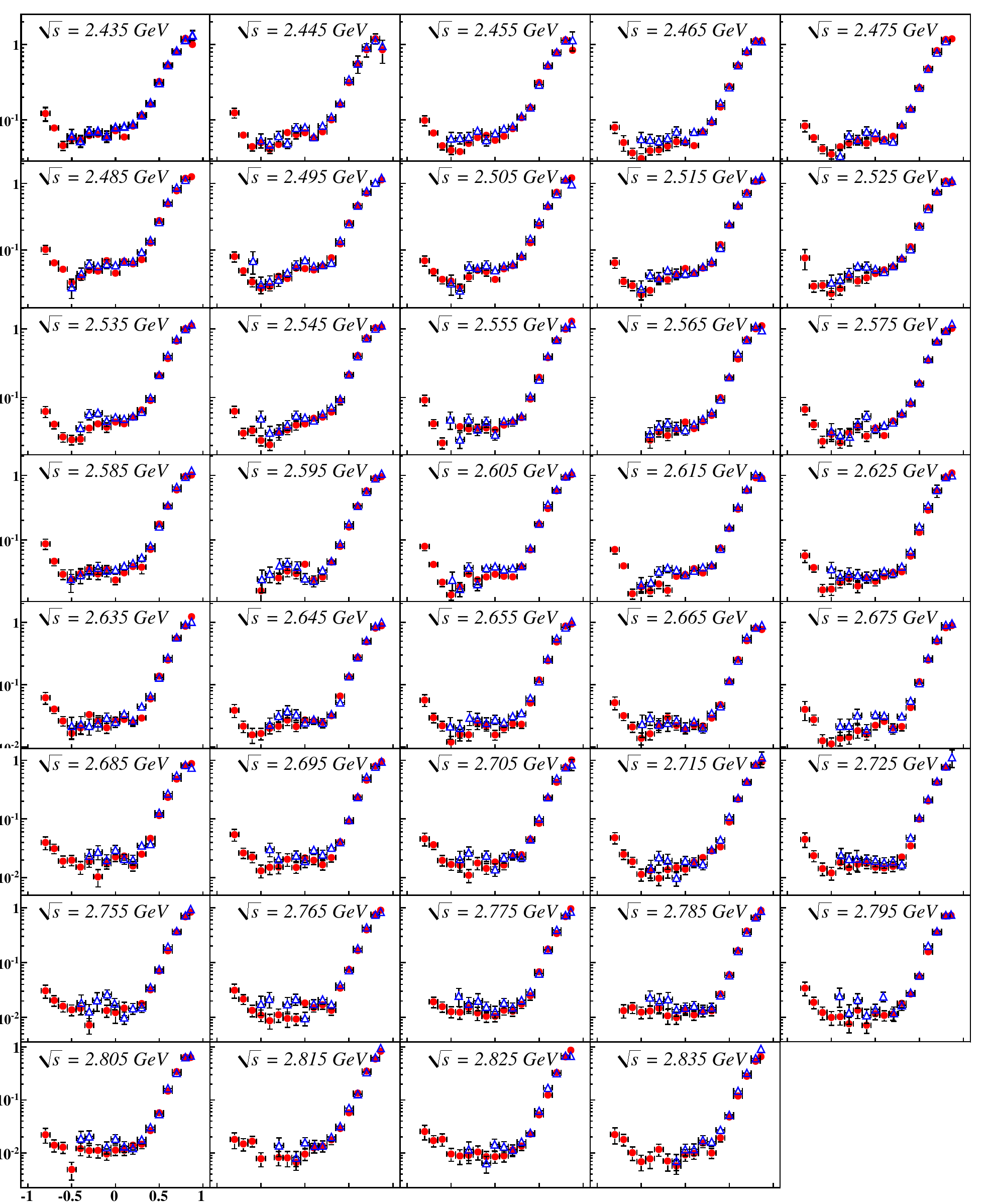}\\
\Large{$\cos\theta_{K}^{c.m.}$}
\caption[]{\label{fig:dsig2}
  (Color On-line)
  $\frac{d\sigma}{d\cos\theta_{K}^{c.m.}}$ ($\mu$b) {\em vs.} $\cos\theta_{K}^{c.m.}$  in bins of $\sqrt{s}$.
Results from the two-track analysis are represented by closed red circles, those of the three-track analysis by open blue triangles.  All error bars represent statistical uncertainties only.  Vertical axes have the same scale in each row, and horizontal axes all have the same scale.
}
\end{figure*}

For differential cross section ($d\sigma/d\cos\theta_{K}^{c.m.}$) and recoil polarization ($P_{\Lambda}$) measurements, each 10-MeV-wide $\sqrt{s}$ bin was further divided into $\cos\theta_{K}^{c.m.}$ bins of width 0.1.
Measurements for these angular bins are reported at the acceptance-weighted bin centroids, the mean of the bin range with non-zero acceptance.
For each topology, in each kinematic ($\sqrt{s}$ ,$\cos\theta_{K}^{c.m.}$) bin, $d\sigma/d\cos\theta_{K}^{c.m.}$ was calculated according to
\begin{eqnarray}
  \frac{d\sigma}{d\cos\theta^{c.m.}_{K}}(\sqrt{s},\cos\theta_{K}^{c.m.}) = \left( \frac{A_{t}}{\mathcal{F}(\sqrt{s})\rho_{t}\ell_{t}N_{A}} \right) \times \nonumber \\
 \;\;\;\; \frac{\mathcal{Y}(\sqrt{s},\theta_{K}^{c.m.})}{(\Delta \cos\theta_{K}^{c.m.})\eta(\sqrt{s},\cos\theta^{c.m.}_{K})},
\end{eqnarray}
where $A_{t}$, $\rho_{t}$, and $\ell_{t}$ are the target atomic weight, density, and length (respectively), $N_{A}$ is Avogadro's constant, $\mathcal{F}(\sqrt{s})$ is the corrected number of photons incident on the target for the given $\sqrt{s}$ bin, $\Delta\cos\theta_{K}^{c.m.}$ is the angular binning width, and $\mathcal{Y}(\sqrt{s},\cos\theta_{K}^{c.m.})$ and $\eta(\sqrt{s},\cos\theta_{K}^{c.m.})$ are the number of data events and acceptance for the given kinematic bin.
Differential cross section results for both two- and three-track analyses are shown in Figs.~\ref{fig:dsig0}-\ref{fig:dsig2}.
The less restrictive two-track analysis presents measurements at more kinematic points.
In total, the two analyses present $d\sigma/d\cos\theta_{K}^{c.m.}$ measurements at 2076 unique kinematic points.
Error bars in these figures represent statistical uncertainties from the numbers of data events and the Monte Carlo acceptance calculation.

Several noteworthy features are present in the data.
For $\sqrt{s}>1.94$~GeV, the forward peak in the data is very prominent, and for $\sqrt{s}>2.4$~GeV the forward peak dominates the differential cross section, suggesting dominance of $t$-channel production mechanisms.
In the $\sqrt{s}$ range from 2.4~GeV to 2.65~GeV, we observe a bump in the differential cross section at intermediate angles, suggestive of $s$-channel production.  
The scale of this feature is small compared to the forward peak, however the feature's presence in several $\sqrt{s}$ bins is quite interesting as production at these energies is considered to be predominantly $t$-channel.
Above $\sqrt{s}\approx1.92$~GeV, a backwards peak is present in the data, and for $\sqrt{s}>2.39$~GeV we observe the forward and backward peaks to be separated by a relatively flat $d\sigma/d\cos\theta_{K}^{c.m.}$.
This backward-angle, high-$\sqrt{s}$ data presents an exciting opportunity to assess $u$-channel contributions to the reaction.

\begin{figure}[]
  \clearpage
  \centering
  \rotatebox{90}{\includegraphics[width=0.35\textwidth]{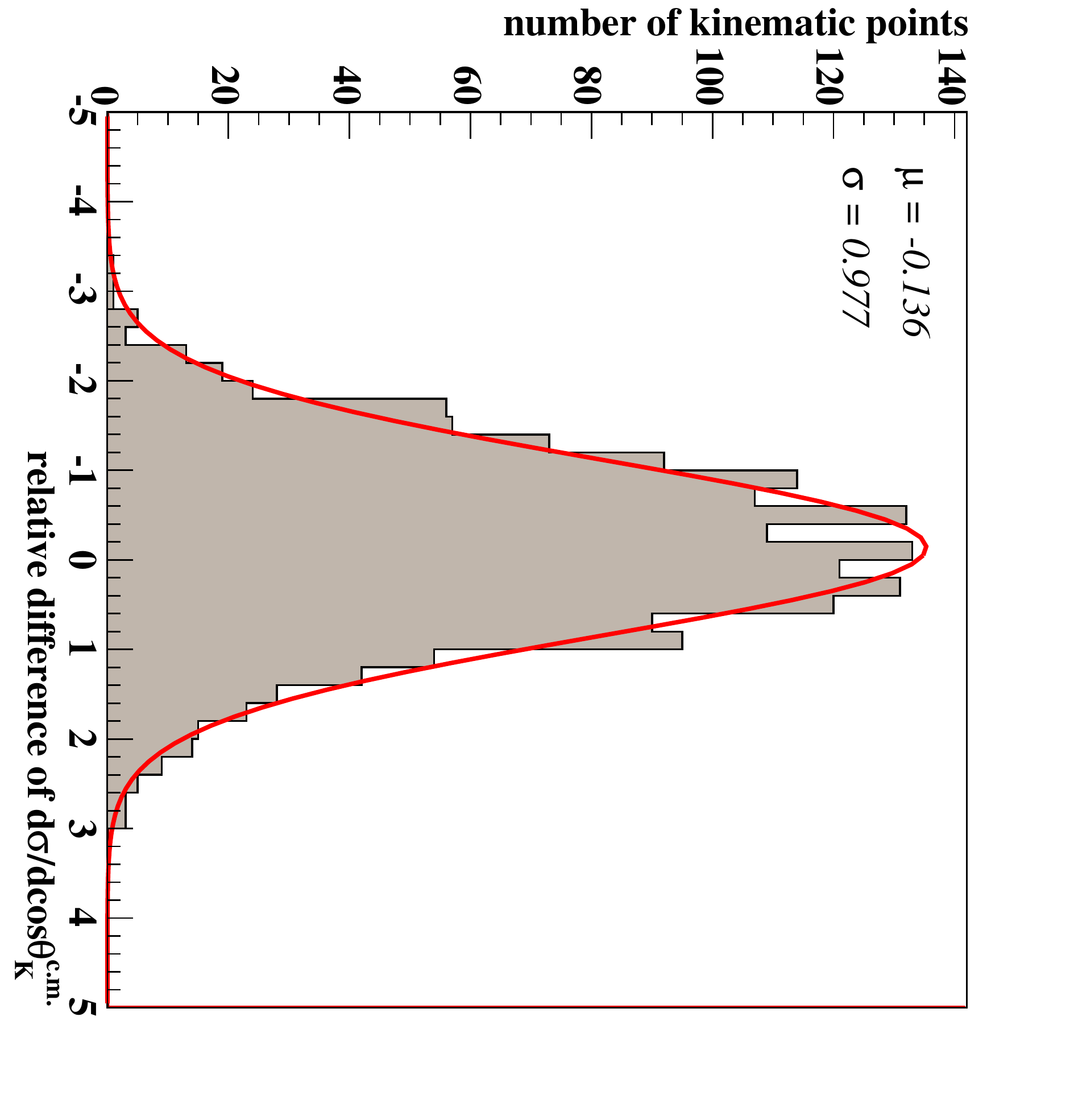}}\\
\caption[]{\label{fig:dsig_dos}
  (Color On-line)
  Relative difference of the two- and three-track $d\sigma/d\cos\theta_{K}^{c.m.}$ results.  
  In red is displayed a fit to a Gaussian function yielding the indicated mean ($\mu$) and width ($\sigma$).  
  See text for details.
}
\end{figure}

Agreement between the two analyses is quantified by the relative difference, $\Delta$, at each kinematic point:
\begin{equation}
  \Delta(\sqrt{s},\cos\theta_{K}^{c.m.}) = \frac{x_{2}-x_{3}}{\sqrt{\sigma_{2}^{2}+\sigma_{3}^{2}+(\overline{x}\sigma_{\eta}(\sqrt{s}))^{2}}},
\end{equation}
where $x_{2(3)}$ and $\sigma_{2(3)}$ are the result and associated statistical uncertainty from the two-track (three-track) analysis, $\overline{x}$ is the average of the two results, and $\sigma_{\eta}(\sqrt{s})$ is the acceptance uncertainty.
This quantity quantifies the difference between the two measurements at a given kinematic point relative to their associated statistical and acceptance uncertainties (e.g., $\Delta=1$ indicates that the difference between two points at a given kinematic is equal to the sum in quadrature of their respective absolute uncertainties).
We find these relative differences to be normally distributed (see Fig. \ref{fig:dsig_dos}) with mean $\mu= -0.136$ and width $\sigma = 0.977$, indicating that the two results show very little systematic offset and are consistent within statistical and acceptance uncertainties.

This level of agreement between the two analyses leads us to produce weighted mean differential cross section results according to
\begin{equation}\label{eq:weighted_mean}
  \overline{x}(\sqrt{s},\cos\theta_{K}^{c.m.}) = ( \displaystyle\sum_{i} \frac{x_{i}}{\sigma_{i}^{2}} ) / ( \displaystyle\sum_{j} \frac{1}{\sigma_{j}^{2}} ),
\end{equation}
where the sums are over the two analyses and $x$ and $\sigma$ represent the measured quantity (here $d\sigma/d\cos\theta_{K}^{c.m.}$) and associated statistical uncertainty.
The statistical uncertainty on these mean values is then given by
\begin{equation}\label{eq:weighted_err}
  \overline{\sigma}^{2}(\sqrt{s},\cos\theta_{K}^{c.m.}) = \left( \displaystyle\sum_{i}1/\sigma_{i}^{2}\right)^{-2} \left( \frac{1}{\sigma_{2}^{2}}+\frac{1}{\sigma_{3}^{2}}+\frac{2\rho}{\sigma_{2}\sigma_{3}} \right),
\end{equation}
where the correlation factor, $\rho=0.28$, is due to the 28\% overlap of the two data samples.
For kinematic points where only a two-track measurement exists, we use it as the mean value and account for the slight offset in the two results by scaling its uncertainty by $1 + |\mu| = 1.136$.

Comparison of these mean $d\sigma/d\cos\theta_{K}^{c.m.}$ results with results of previous experiments are worth comment.
Prior to this analysis, the two highest-statistics studies of $K^{+}\Lambda$ photoproduction (previous CLAS results \cite{bradford} and SAPHIR 2004 \cite{glander}) showed troubling discrepancy.
Most notably, the previous CLAS differential cross sections presented a sizable enhancement at $\sqrt{s}\approx 1.9$~GeV at nearly all production angles, whereas the SAPHIR results showed a monotonically decreasing $d\sigma/d\cos\theta_{K}^{c.m.}$ for $\sqrt{s}>1.75$~GeV and $\cos\theta_{K}^{c.m.}>-0.15$.
Though the magnitude of the discrepancy between these two analyses does not exceed 40\%, the shape discrepancy has a large impact on interpretation of $K^{+}\Lambda$ photoproduction mechanisms.

\begin{figure*}[p]
  \clearpage 
  \centering
\hspace{-1.15cm}\rotatebox{90}{\hspace{8.0cm}\Large{$\frac{d\sigma}{d\cos\theta_{K}^{c.m.}}$ ($\mu$b)}}\includegraphics[width=0.90\textwidth]{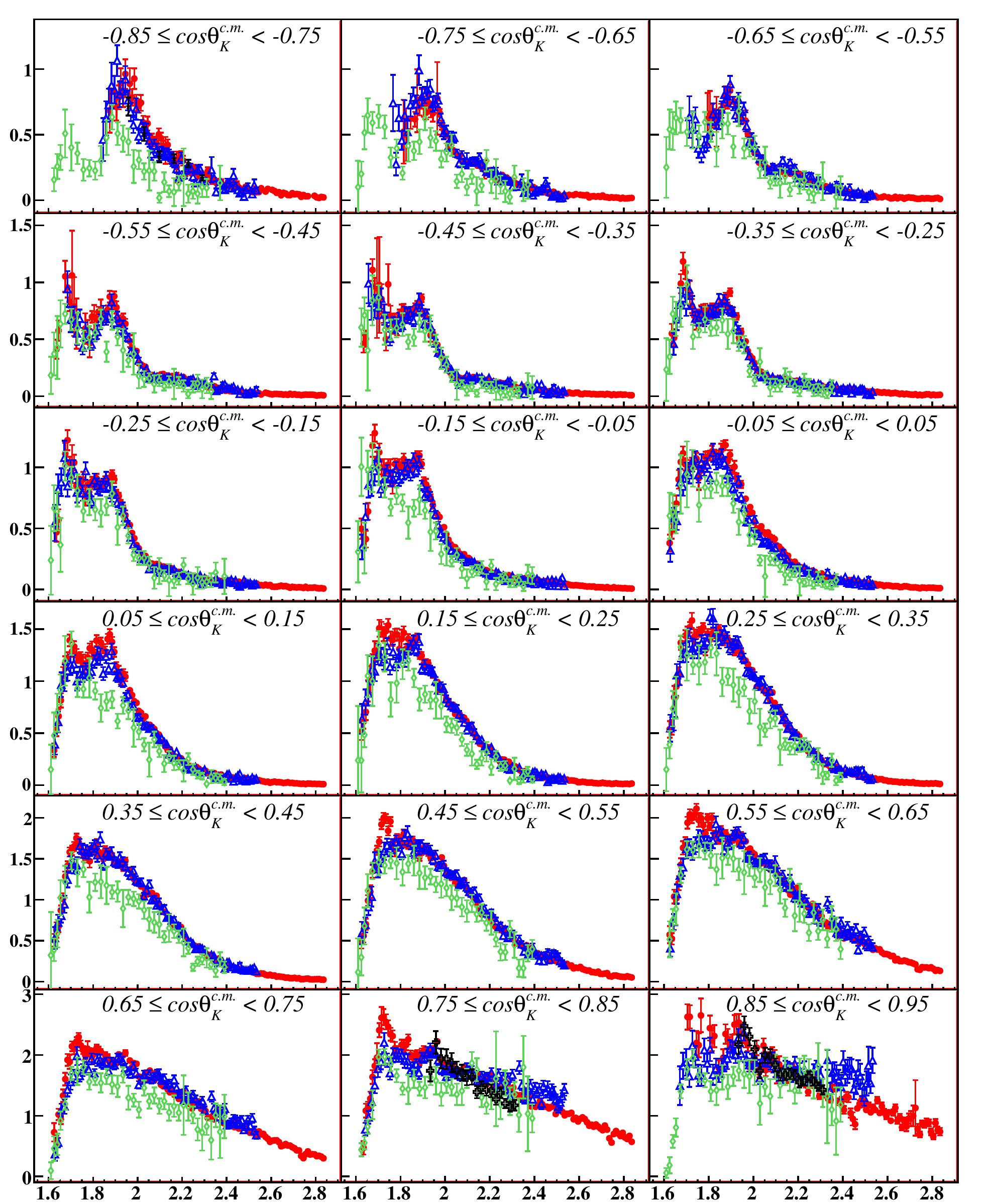}\\
\Large{$\sqrt{s}$ (GeV)}
\caption[]{\label{fig:compare_dsig}
  (Color On-line)
  $d\sigma/d\cos\theta_{K}^{c.m.}$ ($\mu$b) {\em vs.} $\sqrt{s}$ (GeV) in bins of $\cos\theta_{K}^{c.m.}$.
The results of this analysis are shown by closed red circles.
The 2006 CLAS results (Bradford, \textit{et al.} \cite{bradford}) are shown by open blue triangles, 2004 SAPHIR \cite{glander} results are shown by open green diamonds, and the LEPS results \cite{sumihama,hicks} are shown by open black crosses. 
}
\end{figure*}

Fig.~\ref{fig:compare_dsig} shows the results of this analysis plotted with previous high-statistics measurements versus $\sqrt{s}$ in bins of center-of-mass $K^{+}$ production angle.
The new CLAS results confirm the previous CLAS results at most kinematics, most notably at $\sqrt{s}\approx 1.9$~GeV.
These new results also show agreement with forward \cite{sumihama} and backward \cite{hicks} measurements from the LEPS experiment, a very illuminating comparison, as the LEPS results lie at kinematics which are typically at the edges of acceptance for the CLAS and SAPHIR detectors.
We note that these new CLAS results are the most precise to date, and extend the observed $\sqrt{s}$ range for this reaction by $\approx 300$~MeV.

The two CLAS results show excellent agreement in nearly all of the 120 energy bins, but slight systematic discrepancies are present for two specific kinematic regions. 
The first region is that of extreme forward $K^{+}$ production angles ($\cos\theta_{K}^{c.m.}>0.85$).
In this region, the phase space acceptance extrapolation to kaon angles of $0^{\circ}$ used in the earlier CLAS result was probably less accurate than the method used in the present analysis.  
At the extreme forward angle, the two measurements are only marginally consistent within the respective systematic uncertainty estimates.
Also, the CLAS run conditions for the present dataset had the target offset from the center of the detector, thus providing improved forward-angle acceptance.

\begin{figure}[]
  \centering
  \subfigure[]{
    \label{fig:low_w_CLAS_comp:1775}
    \rotatebox{90}{\includegraphics[height=0.53\textwidth]{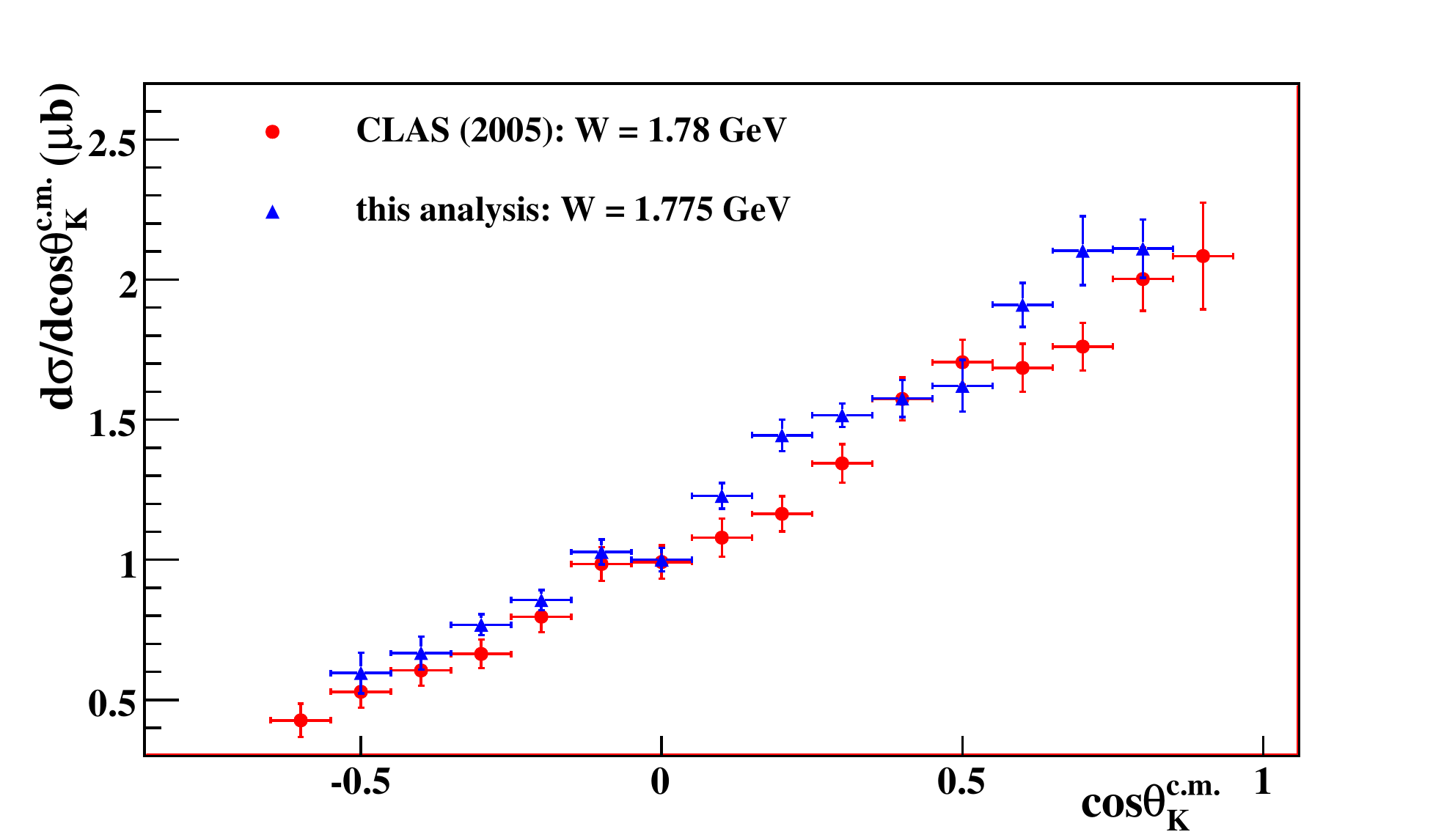}}
  }
  \subfigure[]{
    \label{fig:low_w_CLAS_comp:1745}
    \rotatebox{90}{\includegraphics[height=0.53\textwidth]{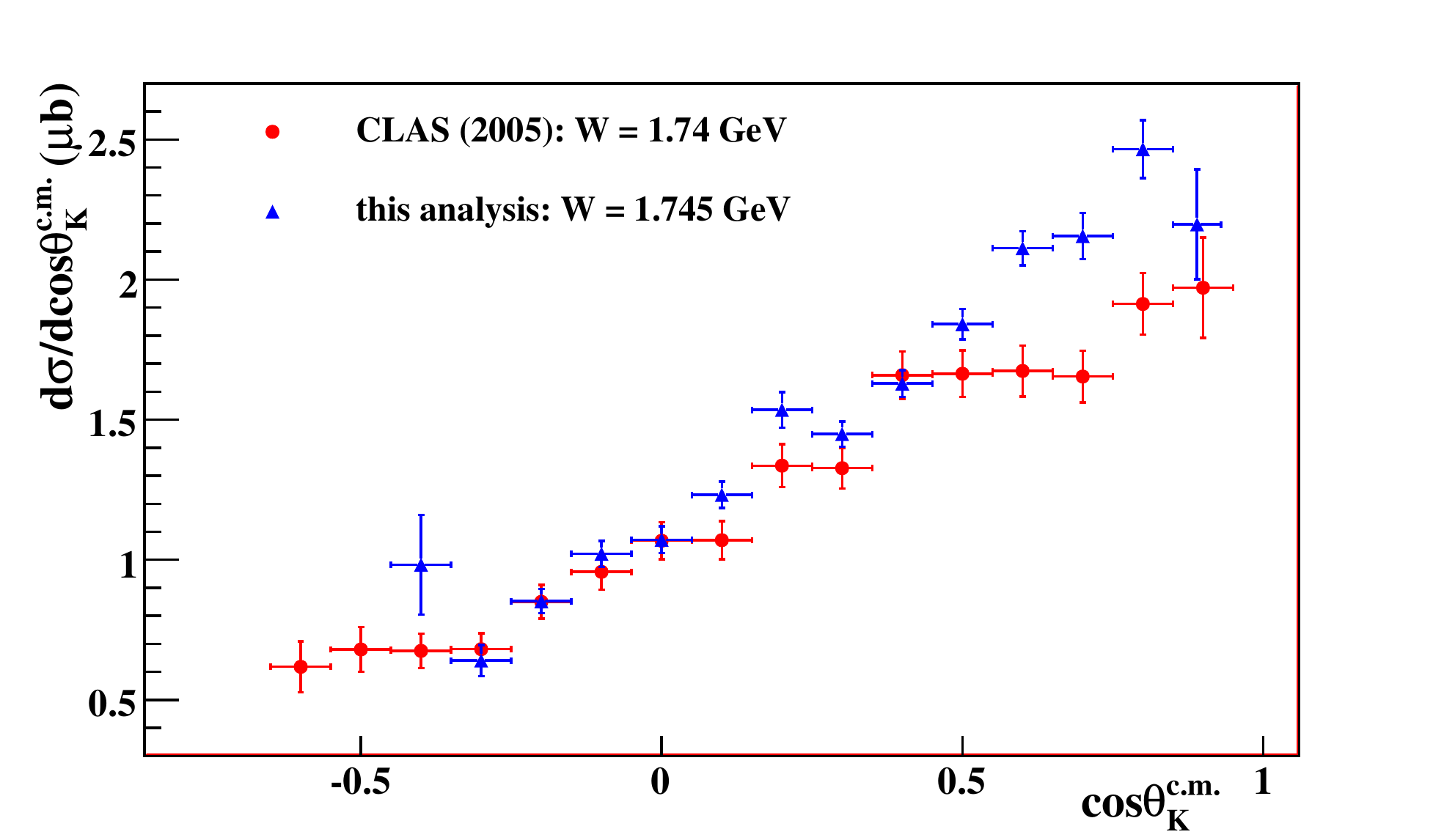}}
  }
  \caption[]{\label{fig:low_w_CLAS_comp}
  (Color On-line)
  $d\sigma/d\cos\theta_{K}^{c.m.}$ {\em vs.} $cos\theta_{K}^{c.m.}$ results from this analysis (blue) and the 2005 CLAS analysis \cite{bradford} (red).
  Fig. \ref{fig:low_w_CLAS_comp:1775} shows results corresponding the $\sqrt{s}=1.775$~GeV bin of this analysis.
  Fig. \ref{fig:low_w_CLAS_comp:1745} shows those of the $\sqrt{s}=1.745$~GeV bin.
  Comparisons in this bin are typical of a four-bin-wide systematic discrepancy between the two datasets.  See text for discussion.
}
\end{figure}

The second region of discrepancy is the four energy bins from $\sqrt{s}=1.715$~GeV to 
$\sqrt{s}=1.745$~GeV. Fig.~\ref{fig:low_w_CLAS_comp:1775} shows the very good agreement 
of the two results in the $\sqrt{s}=1.775$~GeV bin (just outside this region). This bin is an 
example of the typically very good agreement between the two datasets. In 
Fig.~\ref{fig:low_w_CLAS_comp:1745} we present the comparison for the $\sqrt{s}=1.745$~GeV
bin,  that with the largest discrepancy of the four bins. These discrepancies display a dependence 
on production angle, beginning at $\cos\theta_{K}^{c.m.}\approx 0.2$ and continuing to the most 
forward kaon angles. The present results are systematically larger than the previous CLAS results 
at these kinematics, the difference between the two being larger than the results' quoted statistical 
and systematic uncertainties. We have carefully reviewed both analyses, but have been unable to 
identify problems with either. Thus, we are unable to offer unbiased guidance on which data set 
should be preferred for these four energy bins. We can only suggest that in this very narrow energy 
range, the reader exercise care when fitting to the CLAS differential cross section data.

\subsection{$\Lambda$ Recoil Polarization}\label{sec:rec_pol_results}

\begin{figure*}[p]
  \clearpage 
  \centering
\hspace{-1.15cm}\rotatebox{90}{\hspace{8.5cm}\Large{$P_{\Lambda}$}}\includegraphics[width=6.5in]{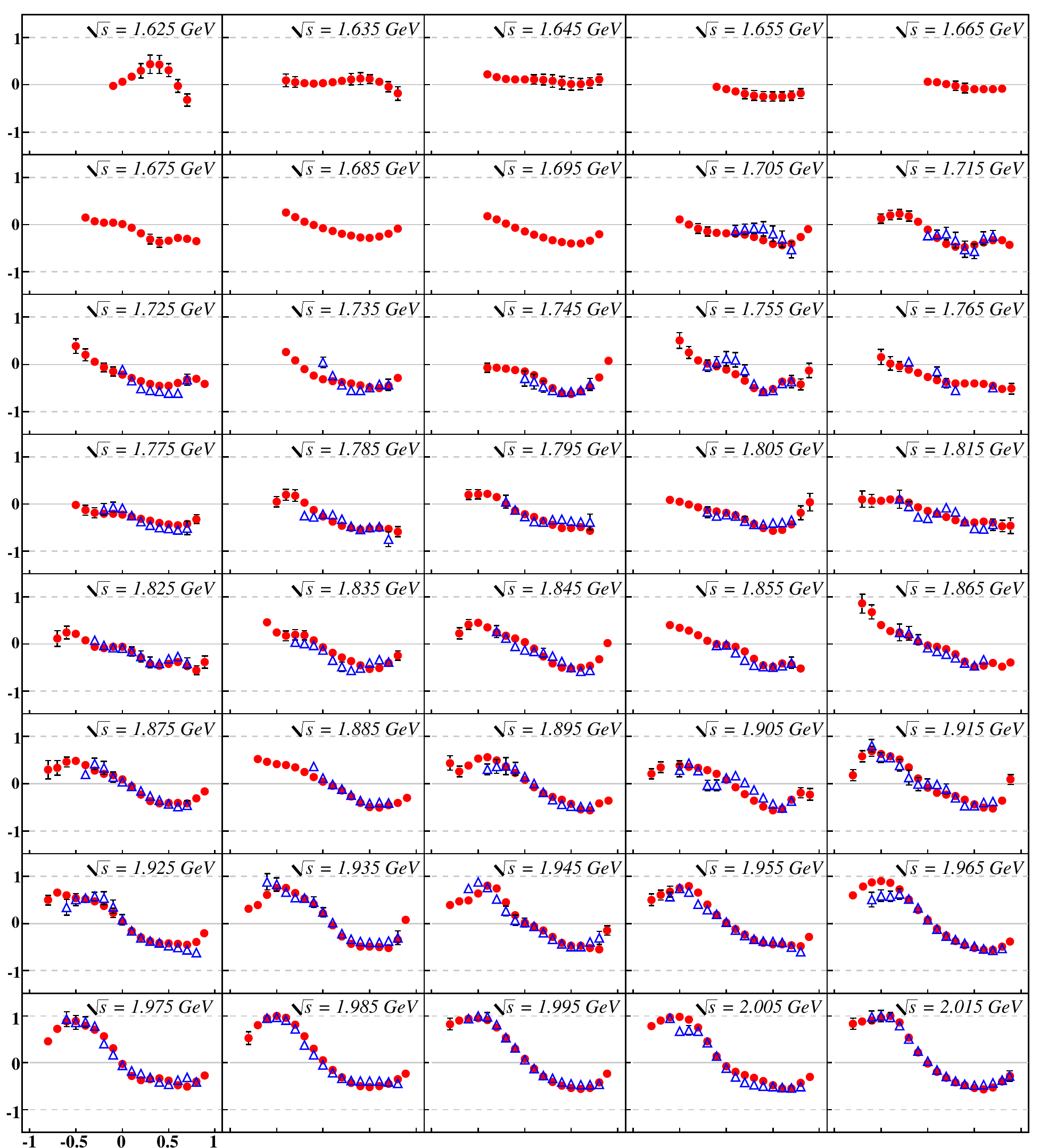}\\
\Large{$\cos\theta_{K}^{c.m.}$}
\caption[]{\label{fig:plam0}
  (Color On-line)
  $P_{\Lambda}$ {\em vs.} $\cos\theta_{K}^{c.m.}$ in bins of $\sqrt{s}$.
Results from the two-track analysis are represented by closed red circles, those of the three-track analysis by open blue triangles.  All error bars represent statistical uncertainties only.  Horizontal and vertical axis scales are common for all plots.  Physical limits on $P_{\Lambda}$ are indicated by dashed horizontal lines.
}
\end{figure*}

\begin{figure*}[p]
  \clearpage 
  \centering
\hspace{-1.15cm}\rotatebox{90}{\hspace{8.5cm}\Large{$P_{\Lambda}$}}\includegraphics[width=6.5in]{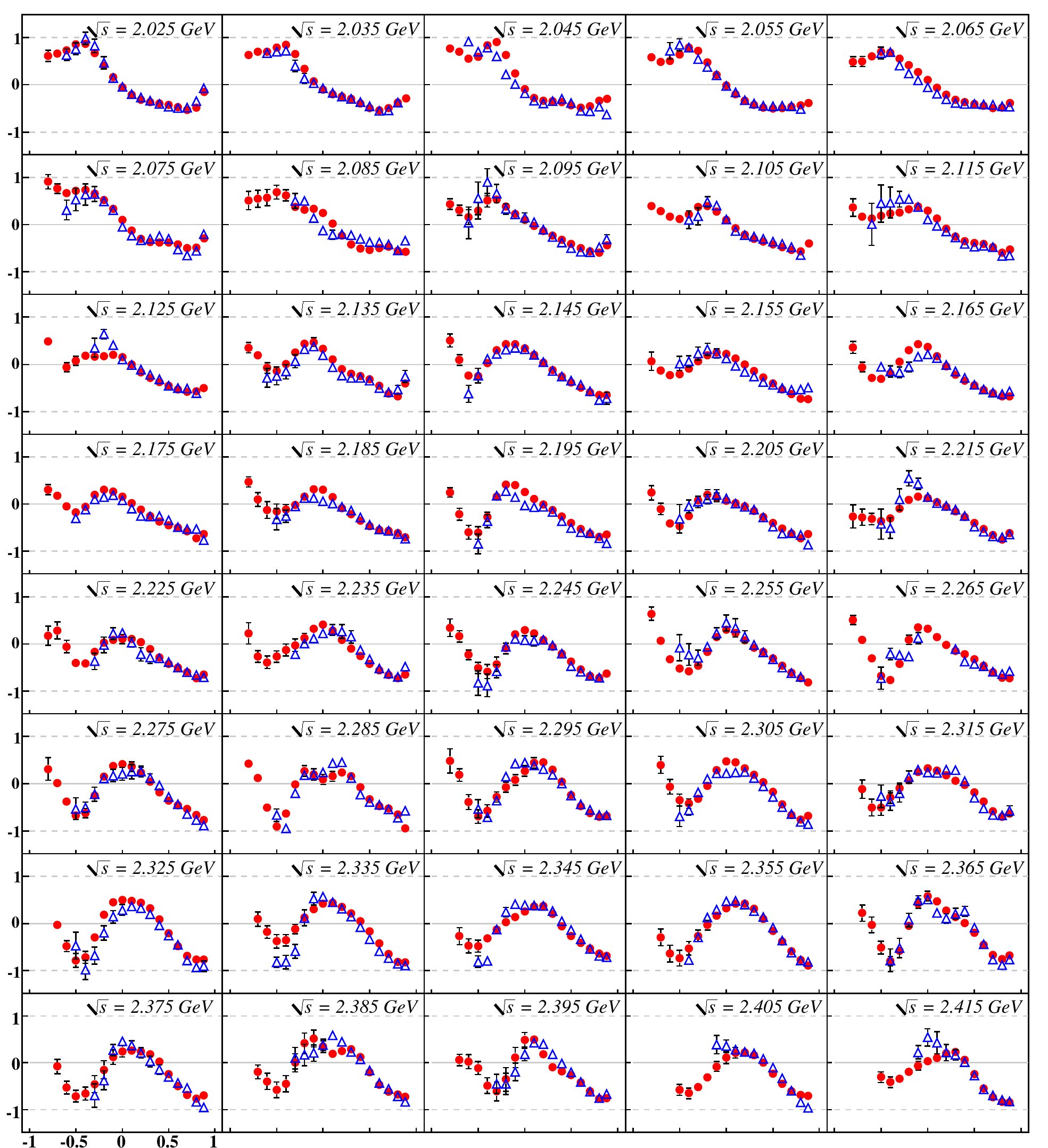}\\
\Large{$\cos\theta_{K}^{c.m.}$}
\caption[]{\label{fig:plam1}
  (Color On-line)
  $P_{\Lambda}$ {\em vs.} $\cos\theta_{K}^{c.m.}$ in bins of $\sqrt{s}$.
Results from the two-track analysis are represented by closed red circles, those of the three-track analysis by open blue triangles.  All error bars represent statistical uncertainties only.  Horizontal and vertical axis scales are common for all plots.  Physical limits on $P_{\Lambda}$ are indicated by dashed horizontal lines.
}
\end{figure*}

\begin{figure*}[p]
  \clearpage 
  \centering
\vspace{-1.0cm}
\hspace{-1.15cm}\rotatebox{90}{\hspace{9.5cm}\Large{$P_{\Lambda}$}}\includegraphics[width=6.5in]{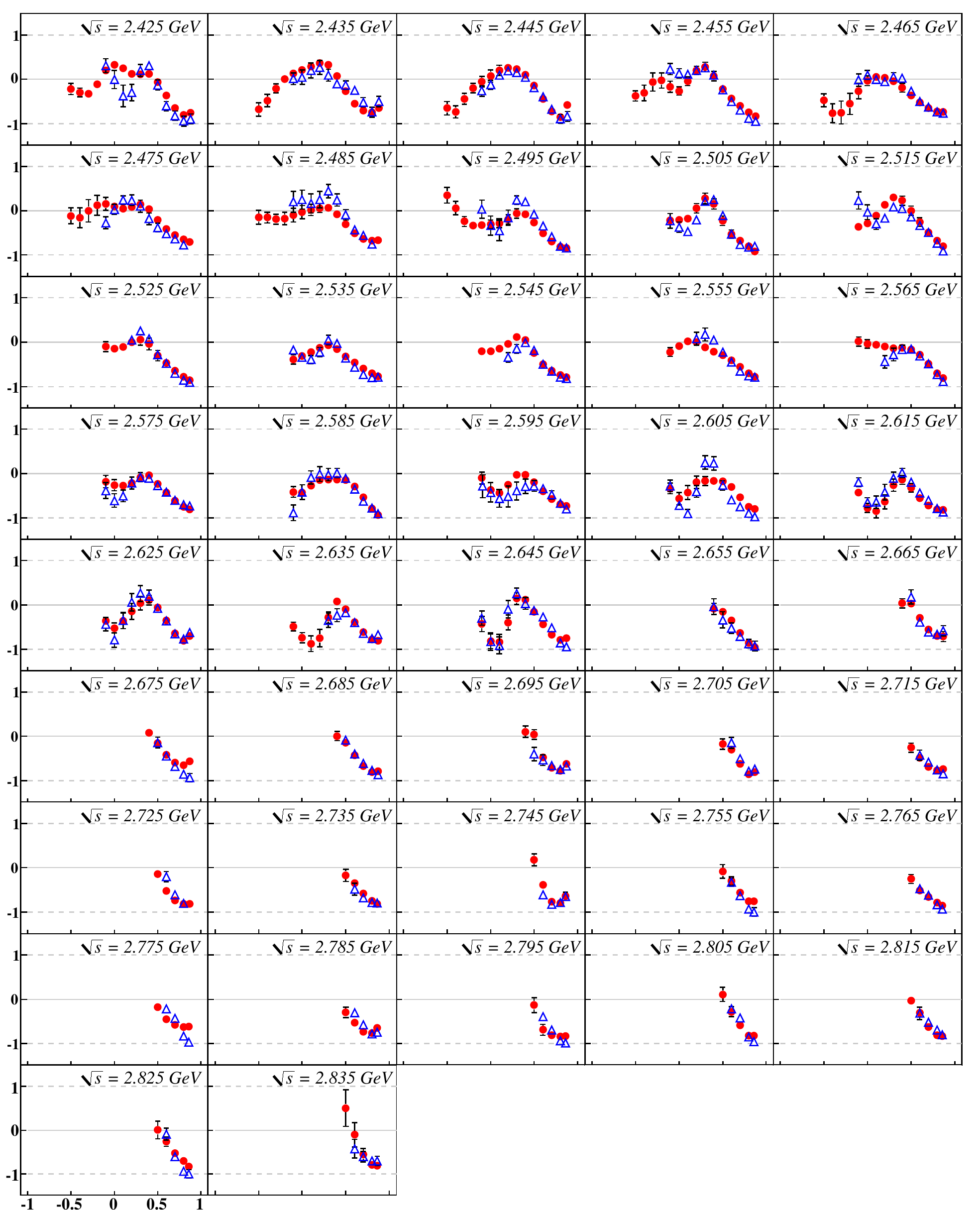}\\
\Large{$\cos\theta_{K}^{c.m.}$}
\caption[]{\label{fig:plam2}
  (Color On-line)
  $P_{\Lambda}$ {\em vs.} $\cos\theta_{K}^{c.m.}$ in bins of $\sqrt{s}$.
Results from the two-track analysis are represented by closed red circles, those of the three-track analysis by open blue triangles.  All error bars represent statistical uncertainties only.  Horizontal and vertical axis scales are common for all plots.  Physical limits on $P_{\Lambda}$ are indicated by dashed horizontal lines.
}
\end{figure*}

$P_{\Lambda}$ results from the two- and three-track analyses are shown in Figs.~\ref{fig:plam0}-\ref{fig:plam2} versus $\cos\theta_{K}^{c.m.}$ in bins of $\sqrt{s}$.
Binning for these results is the same as that used for $d\sigma/d\cos\theta_{K}^{c.m.}$ data.
Error bars in these plots represent statistical uncertainties.
A systematic uncertainty based on acceptance uncertainty discussed in Sec. \ref{section:acc} has been estimated to be 0.05.
In some kinematic areas, differential cross section measurements were possible, however statistics were too low for a reliable $P_{\Lambda}$ measurement.
In all, we present measurements at 1708 kinematic points.

\begin{figure*}[p]
  \clearpage 
  \centering
\hspace{-1.15cm}\rotatebox{90}{\hspace{9.5cm}\LARGE{$P_{\Lambda}$}}\includegraphics[width=0.9\textwidth]{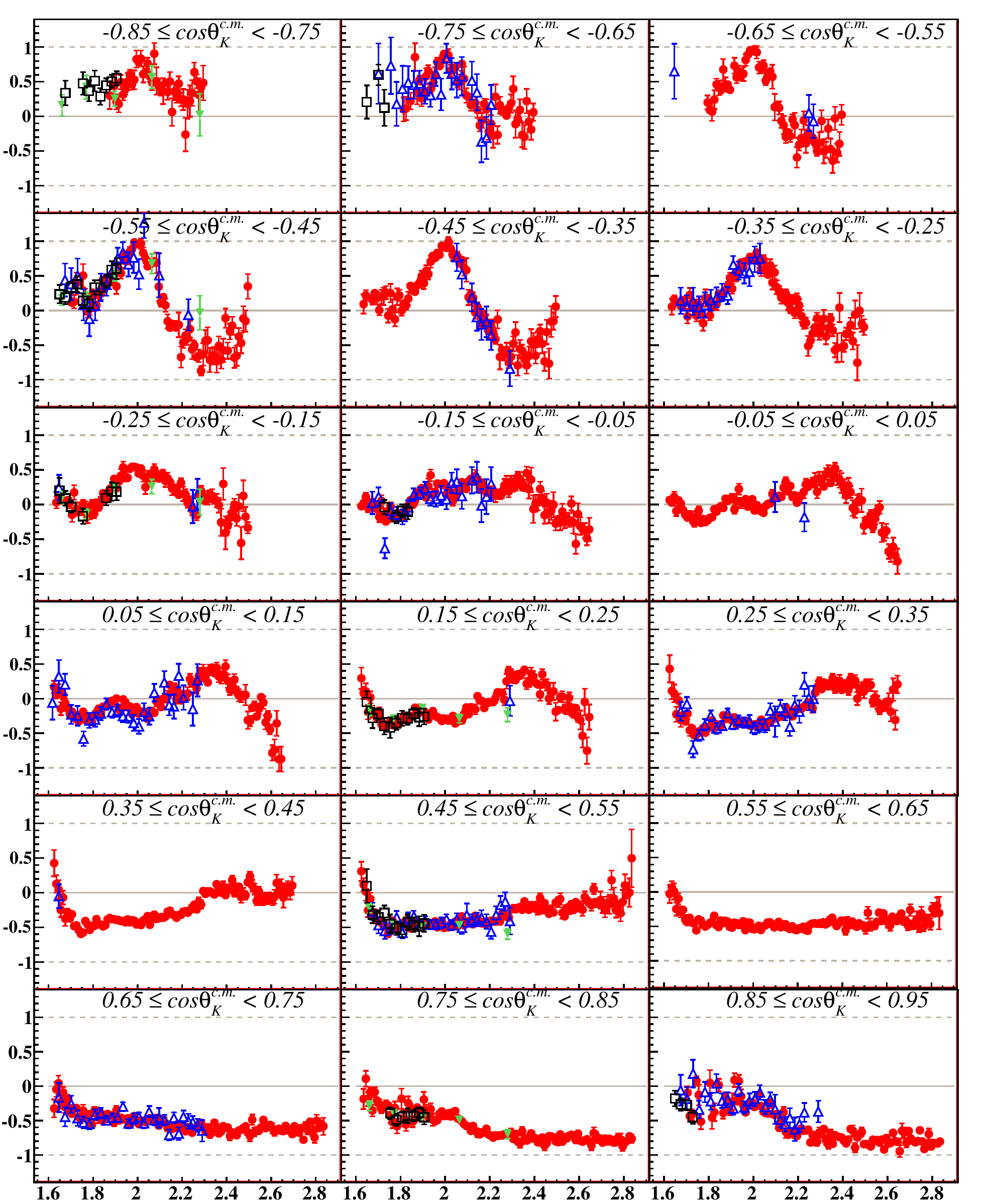}\\
\Large{$\sqrt{s}$ (GeV)}
\caption[]{\label{fig:compare_plam}
  (Color On-line)
  $P_{\Lambda}$ {\em vs.} $\sqrt{s}$ (GeV) in bins of $\cos\theta_{K}^{c.m.}$.  Results of this analysis are represented by red circles, previous CLAS (McNabb, \textit{et al.} \cite{mcnabb}) results by blue triangles, SAPHIR 2004 (Glander, \textit{et al.} \cite{glander}) by green triangles, and GRAAL 2007 (Lleres, \textit{et al.} \cite{lleres}) by black squares.  Physical limits on $P_{\Lambda}$ are indicated by dashed horizontal lines.
}
\end{figure*}

As with the $d\sigma/d\cos\theta^{c.m.}_{K}$ data, we combine the two- and three-track results into a weighted mean result as prescribed by Eqs. (\ref{eq:weighted_mean}) and (\ref{eq:weighted_err}).
Fig.~\ref{fig:compare_plam} shows the mean results plotted with previous high-statistics results from CLAS \cite{mcnabb}, SAPHIR \cite{glander}, and GRAAL \cite{lleres}.
This figure shows the new CLAS measurement's increase in precision and scope, with a nearly 500~MeV increase in $\sqrt{s}$ coverage at forward angles.
The angular resolution of this CLAS measurement is unparalleled by any other measurement.
Comparison between these and existing results presents no systematic discrepancies, and several structures that are hinted at by previous measurements are confirmed by these results.

Several notable structures are present in the $P_{\Lambda}$ data over the $\sqrt{s}$ range from 1.7 to 2.6~GeV.
In the forward direction for $\sqrt{s}>1.9$~GeV, where the reaction is known to be dominated by $t$-channel, the recoil polarization is relatively featureless with respect to $\sqrt{s}$.  
As one looks farther back in production angle, $t$-channel mechanisms become less dominant and undulations in $P_{\Lambda}$ can be seen.
As an example, at backward angles, a region of large positive $\Lambda$ polarization is quite obvious at $\sqrt{s}\approx2.0$~GeV.  
As one looks forward to intermediate angles, the structure remains, but its magnitude is decreased.
Several other bumps are noticeable in $P_{\Lambda}$ at intermediate angles, including those at $\sqrt{s}\approx 2.15$~GeV and $\approx 2.3$~GeV.
We note that for $\sqrt{s}>2.1$~GeV and very forward angles, the recoil polarization remains between $-0.5$ and $-1.0$ indicating a large amount of out-of-production-plane polarization.

\subsection{Model Comparison}

\begin{figure*}[p]
  \clearpage 
  \centering
\includegraphics[width=1.0\textwidth]{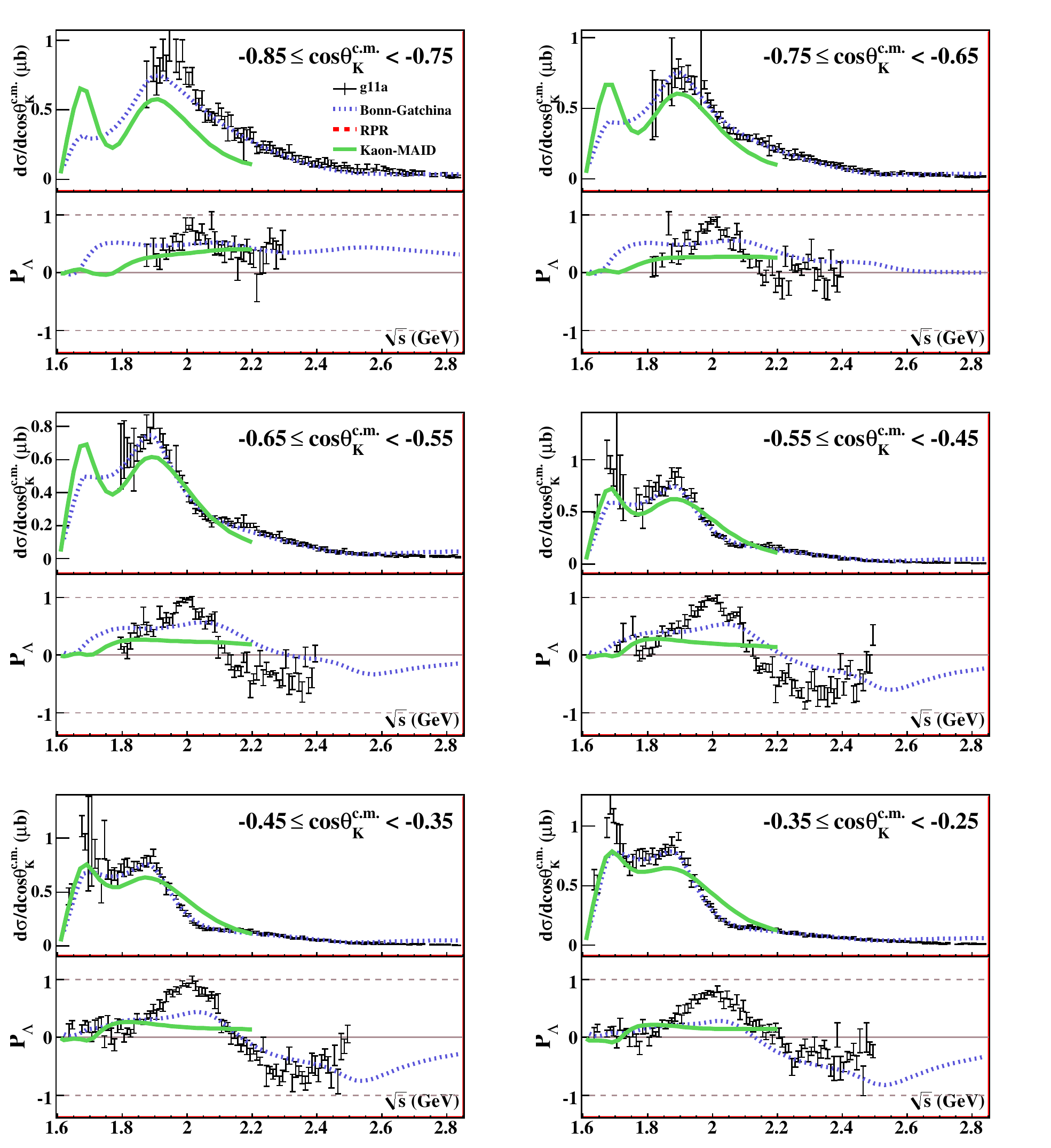}\\
\caption[]{\label{fig:model0}
  (Color On-line) $d\sigma/d\cos\theta_{K}^{c.m.}$ ($\mu$b) and $P_{\Lambda}$ results \textit{vs.} $\sqrt{s}$ (GeV) in bins of $\cos\theta_{K}^{c.m.}$ plotted with several model predictions.  Average data points are given by Eq. (\ref{eq:weighted_mean}).  Model predictions are those of Kaon-MAID \cite{kaon-maid} (solid green line), the Bonn-Gatchina group \cite{sarantsev} (dashed blue line), and the RPR model \cite{corthals} (dashed red line).  See text for commentary.

}
\end{figure*}

\begin{figure*}[p]
  \clearpage 
  \centering
\includegraphics[width=1.0\textwidth]{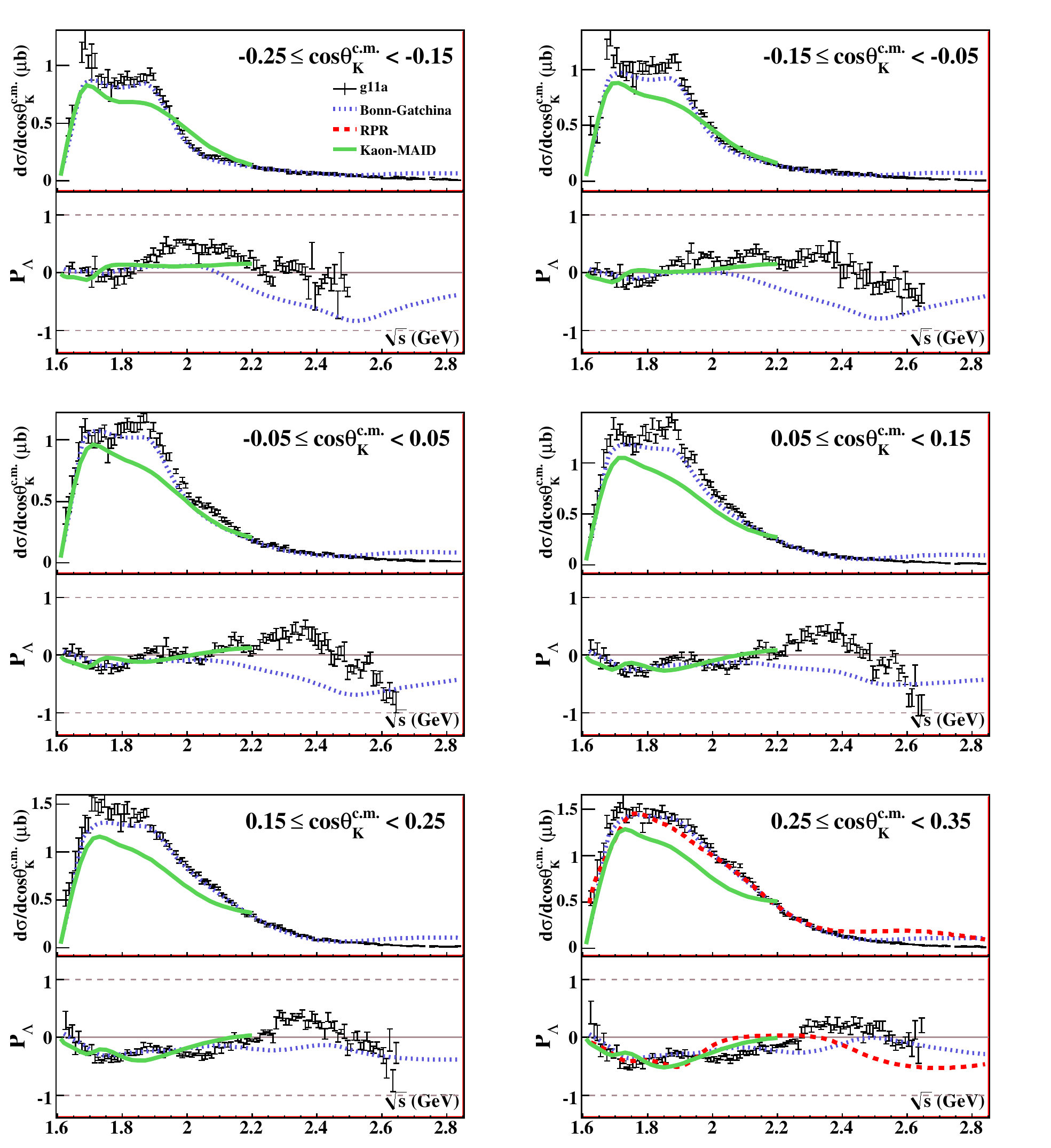}\\
\caption[]{\label{fig:model1}
  (Color On-line) $d\sigma/d\cos\theta_{K}^{c.m.}$ ($\mu$b) and $P_{\Lambda}$ results \textit{vs.} $\sqrt{s}$ (GeV) in bins of $\cos\theta_{K}^{c.m.}$ plotted with several model predictions.  Average data points are given by Eq. (\ref{eq:weighted_mean}).  Model predictions are those of Kaon-MAID \cite{kaon-maid} (solid green line), the Bonn-Gatchina group \cite{sarantsev} (dashed blue line), and the RPR model \cite{corthals} (dashed red line).  See text for commentary.
}
\end{figure*}

\begin{figure*}[p]
  \clearpage 
  \centering
\includegraphics[width=1.0\textwidth]{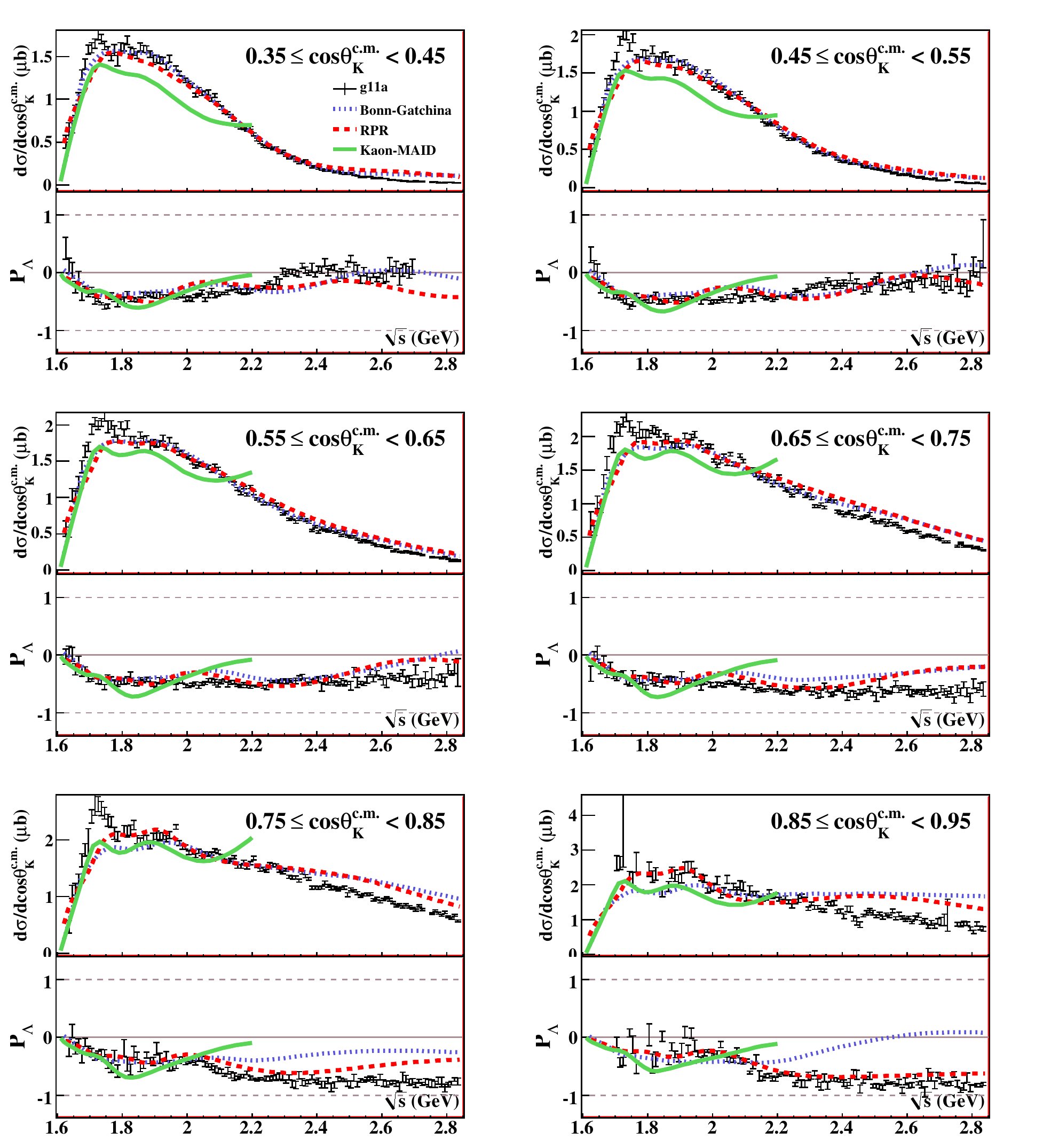}\\
\caption[]{\label{fig:model2}
  (Color On-line) $d\sigma/d\cos\theta_{K}^{c.m.}$ ($\mu$b) and $P_{\Lambda}$ results \textit{vs.} $\sqrt{s}$ (GeV) in bins of $\cos\theta_{K}^{c.m.}$ plotted with several model predictions.  Average data points are given by Eq. (\ref{eq:weighted_mean}).  Model predictions are those of Kaon-MAID \cite{kaon-maid} (solid green line), the Bonn-Gatchina group \cite{sarantsev} (dashed blue line), and the RPR model \cite{corthals} (dashed red line).  See text for commentary.
}
\end{figure*}

For first-order interpretation of features in the data, we compare the average $d\sigma/d\cos\theta_{K}^{c.m.}$ and $P_{\Lambda}$ data (as prescribed by Eq. \ref{eq:weighted_mean}) to the predictions of several contemporary models of $K^{+}$ photoproduction.
Figs.~\ref{fig:model0}-\ref{fig:model2} show the data and predictions of these models \textit{vs.} $\sqrt{s}$ in bins of $\cos\theta_{K}^{c.m.}$.

The Kaon-MAID model \cite{kaon-maid} is an isobar model that treats non-resonant contributions to the channel as $t$-channel exchanges of $K^{+}$, $K^{*}(892)$, and $K_{1}(1270)$ mesons.  
Though the Kaon-MAID model is versatile, the predictions shown here are from a model fit only to SAPHIR data.
Resonant contributions to the channel are attributed to the established $N(1650)$ $S_{11}$, $N(1710)$ $P_{11}$, and $N(1720)$ $P_{13}$ states, as well as a $N(1900)$ $D_{13}$ ``missing'' resonance state necessitated by the enhancement of the differential cross section at $\sqrt{s}\approx 1900$~GeV.
As this model was fit to data of a somewhat limited energy range, predictions are only available below $\sqrt{s}=2200$ MeV.
Because it was tuned to the previous SAPHIR data, scale agreement between the Kaon-MAID model and the present data cannot be expected.
However, conclusions can be drawn from comparisons of specific features of the data and the model.

The second model for comparison is the Regge-Plus-Resonance (RPR) model \cite{corthals} developed by the group at the University of Ghent.
This model treats non-resonant contributions with two Regge-ized $t$-channel exchanges described by a $K^{+}$ Regge trajectory and a $K^{*}$ Regge trajectory (both with rotating phases), an elegant description requiring only three free fit parameters.
As Regge models are often considered valid only for small exchange momenta, the RPR model was tuned only to forward-angle ($\cos\theta_{K}^{c.m.}>0.3$) differential cross section and polarization data from CLAS and previous high-energy data \cite{boyarski}.
Resonant contributions in the RPR model are the $N(1650)$ $S_{11}$, $N(1710)$ $P_{11}$, $N(1720)$ $P_{13}$, and $N(1900)$ $P_{13}$ states, as well as a ``missing'' $D_{13}$ state with a mass of 1900 MeV.
It should not be surprising that this model agrees well with the current $d\sigma/d\cos\theta_{K}^{c.m.}$ results; agreement between these results and the previous CLAS results is satisfactory at most kinematics.

The final model included here is that of the Bonn-Gatchina (BG) group \cite{sarantsev}, which is the result of a large-scale coupled-channel partial-wave analysis of $K^{+}\Lambda$, $K^{+}\Sigma^{0}$, and $K^{0}\Sigma^{+}$, $p\pi^{0}$, $n\pi^{+}$ and $p\eta$ photoproduction data.
It should be noted that the model was constrained to $\gamma p \rightarrow K^{+}\Lambda$ differential cross section, recoil polarization, and beam asymmetry data.
This model employs the operator expansion method, which projects $t$- and $u$-channel amplitudes into $s$-channel partial waves.  
Resonant production in the $K^{+}\Lambda$ channel is represented by significant contributions of the $N(1650)$ $S_{11}$ and $N(1730)$ $P_{13}$ states, as well as two ``newly observed'' $N(1840)$ $P_{11}$ and $N(2170)$ $D_{13}$ states.

Comparison of these models to the new cross section results presents some notable observations.
Though the Kaon-MAID model displays an almost global scale discrepancy, it is evident that the model's treatment of the cross section at $\sqrt{s}\approx 1.9$~GeV (using a ``missing'' $D_{13}$ state) is too weak.
We also note that the Kaon-MAID model overestimates the differential cross section for slightly backward angles and $\sqrt{s}>2.0$~GeV.
The RPR and BG models, as they have been tuned to previous CLAS results match the present results well at most kinematics.
Slight discrepancies exist for the BG model at middle angles and $\sqrt{s}\approx 1.9$~GeV and for the BG and RPR models at forward angles and $\sqrt{s}\approx 1.7$~GeV and $\sqrt{s}>2.4$~GeV.  
At low $\sqrt{s}$, it is possible that these discrepancies can be accounted for by re-tuning the strengths of $s$-channel resonances included.

One feature of the new cross section results that is not reproduced by the models is the slight bump visible at $\cos\theta_{K}^{c.m.} \approx 0.0$ and $\sqrt{s}\approx 2.1$~GeV.  
The PDG lists several $N^{*}$ states with single-star-rated couplings to $K^{+}\Lambda$ near this mass, however, a more systematic analysis of the data should be performed before associating the feature with a given state.

Agreement of these model predictions and the present $P_{\Lambda}$ data is not as good.  
Recall that previous polarization data for this reaction were sparse compared to the present results.
At backward production angles ($\cos\theta_{K}^{c.m.}<-0.15$), we see both the Kaon-MAID and BG models failing to reproduce the large positive polarization of the $\Lambda$ at $\sqrt{s}\approx 2.0$~GeV.
At $\cos\theta_{K}^{c.m.}\approx -0.5$, the models also fail to reproduce the negative $\Lambda$ polarization for $\sqrt{s}>2.2$~GeV.
At intermediate angles ($-0.15\leq\cos\theta_{K}^{c.m.}<0.35$), the BG model reproduces the recoil polarization for $\sqrt{s}<1.85$~GeV, however all three models fail to reproduce the series of bumps in $P_{\Lambda}$ above $\sqrt{s}\approx1.85$~GeV.
As the recoil polarization appears to be very sensitive to the nature of the resonances included, as well as interference between resonances and between resonances and non-resonant mechanisms, these discrepancies could mean that the set of resonances that each of these models employs is either incomplete or incorrect.
It is worth note that for extreme forward angles, only the RPR model seems to accurately describe the recoil polarization (though some further tuning of the model for $0.6\leq\cos\theta_{K}^{c.m.}<0.8$ is called for), lending credence to the Regge-ized meson exchange treatment of non-resonant production.

\section{\label{section:conc}Conclusions}
In conclusion, these CLAS $\gamma p \rightarrow K^{+}\Lambda$ differential cross section and $\Lambda$ recoil polarization results presented here are the most precise to date and offer a significant extension of the observed center-of-mass energy range.
We have presented results from independent analyses of the data and found them to demonstrate satisfying agreement.
These analyses provide $d\sigma/\cos\theta_{K}^{c.m.}$ and $P_{\Lambda}$ measurements at 2076 and 1708 kinematic points, respectively.
The $d\sigma/d\cos\theta_{K}^{c.m.}$ data show satisfying agreement with previous CLAS and LEPS results, while extending the observed $\sqrt{s}$ range by 300~MeV.
These results also provide overwhelming support for the previous CLAS result regarding its discrepancy with SAPHIR results.
The $P_{\Lambda}$ results presented here agree well with all previous results and extend the observed $\sqrt{s}$ range by 500~MeV.
These high-precision measurements show a rich structure in both observables which present an interesting opportunity for interpretation of $K^{+}\Lambda$ photoproduction mechanisms.

\begin{acknowledgments}
The authors thank the staff and administration of the Thomas Jefferson National Accelerator Facility who made this experiment possible.
Thanks also go to Pieter Vancraeyveld of the University of Ghent and Ulrike Thoma and Andrey Sarantsev of the Bonn-Gatchina group for their help in obtaining the model predictions shown in this paper.
This work was supported in part by the U.S. Department of Energy (under grant No. DE-FG02-87ER40315); the National Science Foundation; the Italian Istituto Nazionale di Fisica Nucleare; the French Centre National de la Recherche Scientifique; the French Commissariat \`{a} l'Energie Atomique; an Emmy Noether Grant from the Deutsche Forschungsgemeinschaft; the U.K. Research Council, S.T.F.C.; and the National Research Foundation of Korea.  
The Southeastern Universities Research Association (SURA) operated Jefferson Lab under United States DOE contract DE-AC05-84ER40150 during this work.
\end{acknowledgments}


\end{document}